\documentclass[aps,pre,twocolumn,showpacs]{revtex4}
\usepackage{graphicx}
\usepackage{subfigure}
\usepackage{dcolumn}
\usepackage{bm}
\usepackage{amsmath}

\newcommand{\be}{\begin{equation}}
\newcommand{\ee}{\end{equation}}
\newcommand{\vect}[1]{\ensuremath{\boldsymbol{#1}}}

\begin{document}

\title{Stability of growing vesicles}

\author{Richard G. Morris and Alan J. McKane}

\affiliation{
Theoretical Physics, School of Physics and Astronomy, University of 
Manchester, Manchester M13 9PL, UK}

\begin{abstract} We investigate the stability of growing vesicles using the formalism of nonequilibrium thermodynamics. The vesicles are growing due to the accretion of
	lipids to the bilayer which forms the vesicle membrane. The thermodynamic description is based on the hydrodynamics of a water{/}lipid mixture together with a model of
	the vesicle as a discontinuous system in the sense of linear nonequilibrium thermodynamics.  This formulation allows the forces and fluxes relevant to the dynamic
	stability of the vesicle to be identified. The method is used to analyze the stability of a spherical vesicle against arbitrary axisymmetric perturbations.  It is found
	that there are generically two critical radii at which changes of stability occur. In the case where the perturbation takes the form of a single zonal harmonic, only
	one of these radii is physical and is given by the ratio $2 L_p / L_\gamma$, where $L_p$ is the hydraulic conductivity and $L_\gamma$ is the Onsager coefficient related
	to changes in membrane area due to lipid accretion.  The stability of such perturbations is related to the value of $l$ corresponding to the particular zonal harmonic:
	those with lower $l$ are more unstable than those with higher $l$. Possible extensions of the current work and the need for experimental input are discussed.
\end{abstract}

\pacs{82.20.-w, 05.70.Ln, 87.16.D-} 

\maketitle

\section{Introduction}\label{sec:intro}

The determination of the lowest energy configuration of a vesicle is one of the most widely studied variational problems~\cite{VesReviewSeifert, VesYangLiuXie,
LipowskyNature}.  The vesicle is modeled as a closed two-dimensional surface in three dimensions with a given energy $E_\mathrm{m}$. The subscript $\mathrm{m}$ denotes
``membrane'', since in reality the surface of the vesicle is a membrane in the form of a lipid bilayer~\cite{LifeLuisi}. There are several different models which give
different expressions for $E_\mathrm{m}$~\cite{VesReviewSeifert}. The earliest, and most widely studied, is the Canham-Helfrich-Evans approach~\cite{CanhamJTB, VesHelfrich,
EvansJBiophys}, for which the energy is given by

\begin{equation} 
	E_{\mathrm{m}} = \frac{\kappa}{2} \int \left( 2H - C_{0} \right)^{2}\,\mathrm{d} A, 
		\label{eq:mem_ener} 
\end{equation}

\noindent where $H$ is the mean curvature of the surface with area, $A$, and $\kappa$ and $C_0$ are constants: the so-called bending rigidity and spontaneous curvature
respectively.  The variational calculation might typically consist of finding the shape of the surface which minimizes $E_\mathrm{m}$ for a given vesicle volume $V$ and surface
area $A$.

The popularity of this approach to determining vesicle shape has much to do with the straightforward way it can be posed---not requiring any substantial input regarding
vesicle structure or composition---and the richness of the possible shapes which are found \cite{VesReviewSeifert}. However these reasons are also partly responsible why
the field has been slow to develop: going beyond this description is almost certainly going to involve more of the physics of vesicles and the resulting analysis may not be
so elegant. One of the most obvious drawbacks of the variational studies is that they are static. They give us a snapshot of the shape of the vesicle, but do not tell us
how the shape evolves with time, or the time taken for any new shape to come about.

Several dynamical studies of vesicles have already appeared in the literature~\cite{SvetinaBozicBiophys, SvetinaBozicEurPhysJE, SoleMaciaJTheorBiol, SoleMaciaPhilTransRSB,
VesAJM, BozicSvetinaComment, FanelliMckaneReplyComment, VesRGMDFAJM}. While these preliminary investigations have proved useful in initiating research in this area, all have
been deficient in some way or other. For instance, some of them are not truly dynamic, relying partially on the results of the static analysis~\cite{SvetinaBozicBiophys,
SvetinaBozicEurPhysJE}, while others assumed that vesicle shapes were restricted to spheres or axisymmetric ellipsoids~\cite{VesAJM, VesRGMDFAJM}.  In this paper we describe
a systematic approach to analyzing the stability of growing vesicles.

An obvious question is: what dynamics should be imposed on the system? Following previous treatments, the work presented here uses the formalism of linear nonequilibrium
thermodynamics (LNET). This assumes that that the vesicles are macroscopic~\cite{NonEquilThermDeGrootMazur}. This is reasonable given the size of vesicles, although it is
clear that in some circumstances fluctuations will be important \cite{MWMJ97, WHRS84, US95, VHFS+97, OFPP03}. However it should be noted that similar assumptions were made
deriving the form of the energy (\ref{eq:mem_ener}); it was based on an analogy between the rod-like lipids and nematic liquid crystals, using the methodology introduced by
Frank~\cite{LiquidCrystalEnergyDensityFrank} as motivation. This is a macroscopic \textit{static} description. The equivalent macroscopic \textit{dynamical} description
will involve nematohydrodynamics~\cite{ConservationLiquidCrystalsEricksen, ConstitutiveEqnsLiquidCrystalsLeslie, NematohydrodynamicsHuang}. The approach taken here can
therefore be seen as a natural extension of the static description which leads to (\ref{eq:mem_ener}).

Unlike the static theory, it is necessary to postulate a mechanism which takes the system away from equilibrium, albeit slowly so that LNET holds. One mechanism could be
temperature change, another could be the accretion of lipids onto the surface from the environment. We choose to model the latter mechanism, although we would expect that much of
the formalism constructed will be more widely applicable.

As will become apparent, this study is restricted in two ways.  Firstly, we concentrate on the stability of deformations (as opposed to the full dynamics) and secondly, for
mathematical and presentational simplicity we focus on deformations which take a spherical vesicle to an arbitrary axisymmetric shape.  The outline of the paper is
therefore as follows: in Section \ref{sec:thermodynamics} the thermodynamics of a mixture of point-like constituents (water) and rod-like constituents (lipids) is reviewed with the aim of
identifying the relevant forces and fluxes---and therefore constitutive relations---in a discontinuous LNET description of vesicle dynamics; in Section \ref{sec:diff_geo} the formalism
required to describe the change of shape of the vesicle, is outlined; in Section \ref{sec:stability} these two aspects are brought together to provide a dynamical description of a vesicle
growing due to accretion which is used to study the stability of a spherical vesicle to axisymmetric perturbations. We conclude in Section \ref{sec:conclusions} with a review of the
methodology of our approach and on the prospects for future work. There are three technical appendices one for each of Sections \ref{sec:thermodynamics}, \ref{sec:diff_geo}, and \ref{sec:stability}. 

\section{Thermodynamics}\label{sec:thermodynamics} 

In a previous study~\cite{VesRGMDFAJM} which was restricted to deformations between spheres and ellipsoids, the thermodynamics of vesicle growth was presented in a
straightforward but minimal way.  Here, the aim is to provide a more detailed account.  Some of the necessary theory---LNET, the rheology of nematics and the study of
liquid crystals---is already present in the literature, though for clarity, certain parts are recapitulated (in light of well-known texts) whilst the details of specific
calculations are provided in Appendix \ref{app:discontinuous}.  

\subsection{Linear nonequilibrium thermodynamics}\label{sec:LNET}

The application of LNET to membrane systems has been studied previously \cite{Staverman, DeGrootMazurMichels, LorimerMembrane, BaranowskiMembrane}, however the focus has
been primarily on transport phenomena with the membrane treated as a single discontinuity separating two regions.  In contrast, this paper is concerned with the membrane
itself, and the deformations which occur during the process of growth due to accretion.  It is assumed from the outset that the bilayer is closed (\textit{i.e.}~a vesicle)
and that the surrounding solution is sufficiently dilute that lipids only attach to the surface of the existing bilayer (and do not form other aggregates).
References~\cite{HydrophobicEffect, SelfAssemblyIsrael} have previously considered the aggregation of amphiphiles (lipids) for which the chemical potential of a given
species is taken to be a function of the aggregation number, that is, the number of molecules of the same species in the local neighborhood.  A similar mechanism is
implicitly considered here by assuming that any molecular preference to be part of the bilayer, rather than part of the solution, is controlled by chemical potential
gradients.  

\begin{figure}[h] \centering \includegraphics[scale = 0.8]{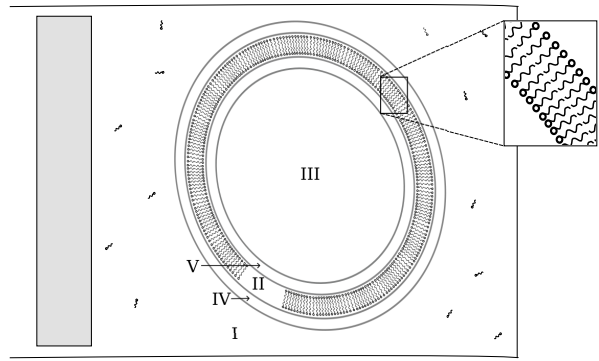} 
	
	\caption{Vesicle system schematic: the system is formed from two distinct phases, dilute water-lipid solution and the lipid bilayer, which are partitioned into three
	regions, the exterior, membrane and the interior, labeled I, II and III, respectively.  
	Thermodynamic variables in regions I and III are taken to be independent of
	position; there are no diffusion flows, viscous flows or chemical potential gradients.   Region II, the membrane, is considered to have reached equilibrium in the sense
	that the molecules are arranged in the usual bilayer configuration (shown in the exploded section); ``tails" pointing inwards and long axis orientated along the surface
	normal.  Changes in the fluid resulting from transport in and out of the membrane are assumed to be confined to very small areas surrounding the membrane boundary,
	these areas are labeled IV and V and are taken to be quasi-stationary, that is, state variables may vary with position but on the timescale of changes experienced in
	regions I, II and III, they are independent of time.  The exterior is taken to behave like a large reservoir while, by contrast, it is assumed that there is no net
	exchange of lipids between the membrane and the interior.}\label{fig:schematic} 

\end{figure}

The usual LNET approach \cite{NonEquilThermDeGrootMazur} is to consider a system sufficiently close to equilibrium that it can be divided into very small sub-systems which are
effectively homogeneous.  Though small, these sub-systems are considered mesoscopic, that is, still large enough to define thermodynamic variables.  It is then possible to
choose a sufficiently large scale on which the variables that characterize each small sub-system form a continuous field.  Each sub-system, and so each point in space on the
larger scale, is taken to obey the Gibbs relation, which can be written in the general form  

\begin{equation} 
	T\mathrm{d}s = \mathrm{d}u + p\mathrm{d}\nu - \left\{\mathrm{d}g\right\}_{T,\ p}, 
	\label{eq:Gibbs} 
\end{equation}

\noindent where all variables are now functions of position and time.  Here, $T$ is the temperature, $p$ is the pressure and following the literature we use $s$, the specific
entropy, given by $S/M$, where $S$ is the entropy and $M$ is the mass.  Similarly, $u = U/M$ is the specific internal energy, $\nu = V/M$ is the specific volume, and $g = u -
Ts + p\nu$ is the specific Gibbs energy.  The subscripted brackets, $\{\ldots\}_{T,\ p}$, are used to indicate that both temperature and pressure are held constant.
Traditionally, the rate of entropy production is then written as a sum of thermodynamic forces and fluxes which are related by constitutive equations.  

For many LNET problems the relevant scale is such that the system effectively comprises a small number of uniform (independent of position) regions.  In these cases,
gradients of thermodynamic variables between regions are taken to be singular.  For such \textit{discontinuous}~\cite{NonEquilThermDeGrootMazur} systems, thermodynamic
forces take the form of differences (rather than gradients) and thermodynamic fluxes become total flows between regions.  In this paper vesicles are treated in such a way;
the membrane is taken to be a separate region of high lipid density---arranged in typical bilayer configuration---with discontinuous transitions to uniform dilute
water-lipid solutions on either side.   

\subsection{Growing vesicles}\label{sec:growingvesicles}

Consider the isolated system described in Fig.~\ref{fig:schematic}; no external forces 
act, no chemical reactions may take place and temperature is taken to be constant 
throughout.  The boundaries of each region are characterized by the outward normal, 
whereby all internal boundaries allow both heat and particle transfer but the external 
system boundary is adiabatic.  Pressure is assumed to be controlled by a piston, shown in 
gray, which allows heat but not particle transfer.  Choosing $\alpha \in \{\textrm{I} - 
\textrm{V}\}$ to label the separate regions of the system the total mass contained in a 
region is given by

\begin{equation} 
	M^\alpha \equiv \sum_k M^{\alpha}_k = \sum_k \int_\alpha \rho_k \mathrm{d} V, 
		\label{eq:M^alpha} 
\end{equation}

\noindent where $\rho_k$ is the partial mass density, $k \in \{l,\ w\}$ is used to label the 
components (lipid and water respectively) and the integral is over volume $V$ of region 
$\alpha$.  With this in place it is possible to introduce the total mass flux of a 
component out of a region

\begin{equation} 
	\frac{\mathrm{d} M^{\alpha}_k}{\mathrm{d} t} \equiv -\int_{\alpha} \rho_k \left( \boldsymbol{v}_k-\boldsymbol{v}^\mathrm{b}\right) \cdot \mathrm{d} \boldsymbol{A}, 
	\quad\forall\ \alpha \in \{\mathrm{I},\ \mathrm{II},\ \mathrm{III}\}, 
		\label{eq:j_k^alpha} 
\end{equation}

\noindent where $\boldsymbol{v}_k$ is the partial velocity of component $k$, 
$\boldsymbol{v}^\mathrm{b}$ is the velocity of the boundary and $\mathrm{d} 
\boldsymbol{A}$ is the area element (aligned along the outward normal) of the surface, 
$A^\alpha$, containing region $\alpha$.  For external boundaries the usual ``no-slip'' 
condition applies: $\boldsymbol{v} = \boldsymbol{v}_k = \boldsymbol{v}^\mathrm{b}$.  
However, in a departure from~\cite{NonEquilThermDeGrootMazur}, internal (permeable) 
boundaries are permitted to move. 

Our approach is to consider a separation of timescales between the changes which occur to 
the membrane and the dynamics of the surrounding fluid.  Both the interior and exterior 
regions are assumed to equilibrate on a timescale much smaller than vesicle growth---that 
is, they are taken to be uniform.  Physically, such an assumption is plausible in the 
light of molecular simulations which indicate that the pressure difference between two 
regions separated by a bilayer are surprisingly large~\cite{OHSO+09}.  There are 
undoubtedly far from equilibrium regimes in which the vesicle is changing quickly enough 
for heterogeneous pressure differences to arise, but these situations are not considered 
in this paper.  Indeed, for the same reason we neglect any flow fields which could arise 
as a result of friction with the moving membrane.  As a result of these assumptions, local 
mass fluxes at the boundary to both regions (interior and exterior) are taken to be 
independent of position.  By contrast, the membrane, however, is not uniform.  As indicated 
in Fig.~\ref{fig:schematic}, molecules are assumed to be orientated in a bilayer fashion, 
so that for any shape other than a sphere, the local configuration of lipids (e.g.~molecular 
splay) has an angular dependence.  We make the assumption that such differences do not 
mechanically affect the flow of mass, either water or lipids, into or out of the membrane.  
Taking partial velocities to be in the direction of the outward normal, local mass fluxes, 
given by $\rho_k \left( v_k-v^\mathrm{b}\right)$ are also assumed to be constant at the 
boundary to the membrane region.

The relative configuration of lipids in the membrane is, however, still considered 
important thermodynamically.  Indeed, for such a simplified description of vesicles---with 
a uniform interior and exterior, driven by mass fluxes which do not vary at different 
points on the membrane---it seems reasonable to expect that any dynamical behavior (or 
shape change) will involve averaging some thermodynamic quantity over the membrane.  For 
example, two vesicles of different shapes but equivalent \textit{average} molecular splay 
are anticipated to undergo dynamics driven by the same total flows between interior, 
exterior and membrane regions.  With this in mind, confining the details to Appendix 
\ref{app:discontinuous}, we find that the total entropy produced in the system is given by

\begin{equation} 
	\sigma_{\mathrm{tot}} = \frac{1}{T}\sum_k \sum_{\alpha=\mathrm{II}}^{\mathrm{III}} 
	\left( \bar{\mu}_k^{\mathrm{I}} - \bar{\mu}_k^{\alpha} \right) \frac{\mathrm{d} M^{\alpha}_k}{\mathrm{d} t},
	\label{eq:sigma_tot_main} 
\end{equation}

\noindent where $\mu_k$ is the chemical potential of component $k$ and a bar above a variable denotes the average value taken over the \textit{boundary} to a region, in the
sense that

\begin{equation}
	\bar{x}^\alpha \equiv \frac{1}{A^\alpha} \int_\alpha x \mathrm{d} A,
	\label{eq:bar_x}
\end{equation}

\noindent for some thermodynamic variable $x$.  Immediately it is clear that $\bar{\mu}^\alpha_k = \mu_k$ for $\alpha \in\{\mathrm{I},\ \mathrm{III}\}$.  (The average over
the boundary to region II is addressed below).  It should be noted that in order to reach the result (\ref{eq:sigma_tot_main}), one important assumption has been made: that
the entropy produced due to the re-alignment of molecules as the vesicle grows can be neglected.  This is plausible in the context of a stability analysis of deformations;
any such entropy produced as a result of a small perturbation in the shape is proportional to thermodynamic forces (such as pressure and chemical potential gradients)
\textit{within} the membrane.  These gradients are considered negligible on the scale of pressure/chemical potential differences \textit{across} the membrane.

One might now ask: what role is played by the energy which arises from lipid interactions in the membrane?  In order to answer this, we suppose that the internal energy
averaged over the boundary to region II (the membrane) is well defined thermodynamically, in the sense that $\bar{u}^\mathrm{II} = \bar{u}^\mathrm{II} (\bar{s}^\mathrm{II},
\bar{\nu}^\mathrm{II}, \{\bar{c}_k^\mathrm{II} \}, \bar{\psi}^\mathrm{II})$.  Here, $u$, $s$ and $\nu$, are defined earlier, $c_k = M_k / M$ is the concentration (of
component $k$) and  $\psi = \Psi / M$ is introduced as a specific extensive variable characterizing orientation dependent---in this case, amphiphilic---interactions.
Dropping superscripts for simplicity, this implies   

\begin{equation} 
	\mathrm{d}\bar{u} = T\mathrm{d}\bar{s} - p\mathrm{d}\bar{\nu} + \sum_k\bar{\mu}_k\mathrm{d}\bar{c}_k + \bar{\mu}_\psi\mathrm{d}\bar{\psi}, 
		\label{eq:bar_u}
\end{equation}

\noindent where $\mu_k$ is the chemical potential of component $k$ and

\begin{equation} 
	\bar{\mu}_\psi = \left( \frac{\partial \bar{u}}{\partial \bar{\psi}} \right)_{T,\ p,\ \{\bar{c}_k\}}. 
	\label{eq:mu_Gamma} 
\end{equation}

\noindent Note that for this system $\bar{T}=T$ and $\bar{p} = p$.  Assuming that the average specific internal energy is homogeneous and of first order in the masses of each
component implies that the average specific Gibbs energy may be written as

\begin{equation} 
	\bar{g} = \sum_k\bar{\mu}_k \bar{c}_k + \bar{\mu}_\psi \bar{\psi}, 
	\label{eq:gapp} 
\end{equation}

\noindent that is, a sum of the usual free energy of a system of point-like constituents---arising due to the concentrations of components---and another term relating to the
nematic nature of the molecules.  As previously mentioned, we use the simplest macroscopic model of energy associated with the orientation of lipids in a bilayer: that
attributed to Canham, Evans and Helfrich.  Indeed, ignoring concentration dependent terms (and re-introducing superscripts for clarity), comparison with (\ref{eq:mem_ener})
leads to the following identifications

\begin{equation} 
	\bar{\mu}_\psi^\mathrm{II} = \kappa\ \mathrm{and}\ \bar{\psi}^\mathrm{II} = \frac{1}{2 l M A^\mathrm{m}}\int_\mathrm{m}\left( 2H - C_0 \right)^2\mathrm{d} A, 
	\label{eq:mu_gamma_and_gamma}
\end{equation}

\noindent as $\kappa$ does not scale with the system size (\textit{i.e.}~it is intensive).  In the above the membrane is taken to be of constant thickness $l$, where the
factor preceding the integral comes from the fact that the Canham-Helfrich-Evans model is an integral over energy \textit{per unit area} \cite{PetrovBivas}.  Note also that
the factor of $M$ comes from the fact that $\psi$ is a specific, or per unit mass, quantity.  Here, in order to be consistent with membrane model literature (see for
example the early sections of \cite{VesReviewSeifert} or \cite{PetrovBivas}), the previously defined ``bar averaging'' taken over the boundary to the  membrane has been
replaced by an average over a surface bisecting the two layers of lipids, $A^\mathrm{m}$, the so-called neutral surface~\cite{PetrovBivas}.  Furthermore, this approach is
extended to all thermodynamic variables: an average taken over the exterior of the membrane is the same as an average taken over the neutral surface.  

With this in place, it is now possible to examine how the chemical potential in (\ref{eq:sigma_tot_main}) depends on the other thermodynamic variables.  Assuming the form
$\bar{\mu}_k = \bar{\mu}_k(T, p,\{\bar{c}_k\}, \kappa)$ we see that  

\begin{equation} 
	\mathrm{d}\bar{\mu}_k = \bar{\nu}_k \mathrm{d}p + \bar{\psi}_k \mathrm{d}\kappa + \{\mathrm{d}\bar{\mu}_k\}_{T,\ p,\ \kappa}, 
		\label{eq:dmu_k} 
\end{equation}

\noindent where temperature has been held constant and partial specific quantities $\bar{\nu}_k$ and $\bar{\gamma}_k$ are defined in the following way

\begin{equation} 
	\bar{\nu}_k\equiv\left( \frac{\partial \bar{\mu}_k}{\partial p} \right)_{T,\ \{\bar{c}_{i\neq k}\},\ \kappa}, 
	\label{eq:nu_k} 
\end{equation}

\begin{equation} 
	\bar{\psi}_k\equiv\left( \frac{\partial \bar{\mu}_k}{\partial \kappa} \right)_{T,\ p,\ \{\bar{c}_{i\neq k}\}}.  
	\label{eq:-s_k} 
\end{equation}

\noindent For a discontinuous system such as ours it is useful to first define the difference notation (for some thermodynamic variable $x$)

\begin{equation} 
	\Delta_{\alpha,\ \beta\ } \bar{x} \equiv \bar{x}^{\alpha} - \bar{x}^{\beta},\quad \forall\ \alpha\in\{\mathrm{I,\ II,\ III}\}.  
	\label{eq:Delta_alpha_beta} 
\end{equation}

\noindent Once again following \cite{NonEquilThermDeGrootMazur} it is now possible to write an equivalent expression to (\ref{eq:dmu_k}) for small finite differences
between regions:

\begin{equation} 
	\begin{split} 
		\Delta_{\alpha,\ \beta\ }\bar{\mu}_k =& \bar{\nu}_k^\beta \left( \Delta_{\alpha,\ \beta\ }p \right) + \bar{\psi}_k^\beta \left( \Delta_{\alpha,\ \beta\ }\kappa \right) \\ 
		& + \{\Delta_{\alpha,\ \beta\ }\bar{\mu}_k\}_{T,\ p,\ \kappa}.  
		\label{eq:Deltamu_k} 
	\end{split} 
\end{equation}

\noindent Restating (\ref{eq:sigma_tot_main}) using the above difference notation and then substituting for (\ref{eq:Deltamu_k}) it can be seen that, for small deformations,
the rate of entropy produced will have contributions from pressure differences, energy differences due to membrane deformation and chemical potential differences.  The
resultant expression may then be simplified by applying a number of assumptions which are valid for the type of vesicle system discussed here.  Firstly, the mass of lipids
accreting to the membrane from the interior is negligible.  That is, in contrast to the exterior, the interior is not treated as a reservoir.  Secondly, the membrane
thickness, $l$, is considered small on the scale of the system.  This permits us to write the volume of the membrane, to first order in small parameter $l$, as the area of
the surface which bisects the membrane multiplied by the thickness (i.e.~$V^\mathrm{II} = lA^\mathrm{m} + \mathcal{O}(l^2)$).  Finally, we assume that both $\kappa$, the bending rigidity, and
$C_0$ the spontaneous curvature, remain constant.  For such a case, the average density of lipids in the membrane and the average ratio of lipids between the inner and outer
layers must be unchanged.  Therefore each new unit of mass added to the membrane is assumed to increase the area of the surface which bisects the membrane by a constant
factor.  The manipulations are left to Appendix \ref{app:discontinuous}, where it is shown that  

\begin{equation} 
	\begin{split} 
		T\sigma_{\mathrm{tot}} =&  \Delta p \left\{\frac{\mathrm{d} V}{\mathrm{d} t}\right\}_{T,\ p,\ \kappa} - \left\{\frac{\mathrm{d} E_\mathrm{m}}{\mathrm{d} t}
		\right\}_{T,\ p,\ \kappa} \\ 		
		& + \gamma\left\{\frac{\mathrm{d} A}{\mathrm{d} t}\right\}_{T,\ p,\ \kappa},
			\label{eq:sigma_tot_no_superscripts}
	\end{split} 
\end{equation}

\noindent where $\gamma$ is the surface tension \footnote{Note that in a previous paper~\cite{VesRGMDFAJM}, surface tension was given by the symbol $\sigma$.  Here the
alternative convention of using $\gamma$ is adopted as $\sigma$ represents the rate of entropy production.} and $\Delta p = p^\mathrm{I} - p^\mathrm{III}$ is the pressure
difference between the exterior and interior.  Here, surface area $A$, volume enclosed $V$ and bending energy $E_\mathrm{m}$ are defined in relation to a single mathematical
surface taken to bisect the two lipid monolayers.

It is worthwhile noting here that $\gamma$ corresponds to the ``interfacial free energy'' 
in the sense defined in~\cite{OFPP03}.  That is, the change in free energy which results 
from increasing (decreasing) the surface by adding (removing) molecules at constant 
density---\textit{cf}.~Eq.~(\ref{eq:dM_dt=adA_dt}).  This contrasts with increasing the 
surface area by reducing the density of a fixed number of molecules: the resultant change 
in free energy from such an approach is referred to as the ``elastic free energy''.  Here, 
as with the majority of studies to date, the effects of elastic free energy are neglected 
and the lipid bilayer is assumed to be effectively incompressible due to the separation of 
energy scales between stretching and bending energies~\cite{VesReviewSeifert}.  As pointed 
out in~\cite{OFPP03}, thermal fluctuations are thought to be more important to studies of 
elastic free energy rather than the interfacial energy considered here.

The result (\ref{eq:sigma_tot_no_superscripts}) allows us to identify the forces and fluxes for this nonequilibrium system, and this is discussed further in 
Section \ref{Effective_pressure_and_surface_tension}. In our previous work \cite{VesRGMDFAJM} these were identified largely through physical arguments and from the work of
Kedem and Katchalsky \cite{KedemKatchalBiophys, KedemKatchalFaraday}. We did use simple thermodynamic relations, but only to show that the term involving the membrane
energy could be absorbed into effective forces---an analysis which is recapitulated in Section \ref{Effective_pressure_and_surface_tension}. The thermodynamic analysis 
is this paper is far more extensive, and as such is not easy to compare with that presented in \cite{VesRGMDFAJM}. This is especially true of the identification of the 
various contributions to the entropy. In \cite{VesRGMDFAJM} we considered a fluid separated into two homogeneous regions and wrote down the sum of entropy changes associated 
with each region.  However, here we consider the membrane as a separate region. With proper consideration given to conservation laws, such a sum is equal in size but opposite 
in sign to the entropy produced by the system. As long as care is taken to give the correct interpretation to the various contributions, both treatments agree, but with
the present paper giving a mathematical justification to the form of the dynamical equations used in \cite{VesRGMDFAJM} that was not present in that previous discussion.

\subsection{Effective pressure and surface tension}
\label{Effective_pressure_and_surface_tension}

This paper examines the stability of deformations away from a sphere \textit{i.e.}~does a deformation grow or decay?  In this context it is possible to
simplify (\ref{eq:sigma_tot_no_superscripts}) by appealing once again to the rationale used earlier when neglecting the entropy contributions from molecular re-alignments
within the membrane.  It is assumed that the rate of change in energy due to small membrane deformations is a function of only two time-dependent variables: the surface
area and volume of the membrane.  That is, on small timescales, changes in the energy of the membrane are dominated by the addition of lipids to the surface or by changes
in the pressure difference across the membrane.  This point will be re-examined in more detail in Section \ref{sec:diff_geo}.  For now, taking  $E_\mathrm{m} = E_\mathrm{m}(V,\ A)$ the rate
of entropy production can be written in terms of an \textit{effective} pressure difference and \textit{effective} surface tension. That is

\begin{equation}
\begin{split}
	\left\{ \frac{ \mathrm{d} E_\mathrm{m} }{ \mathrm{d} t } \right\}_{T,\ p,\ \kappa} =& \left( \frac{\partial  E_\mathrm{m}}{\partial V}\right)_A \left\{\frac{\mathrm{d} V}{\mathrm{d} t} \right\}_{T,\ p,\ \kappa} \\
	&+ \left(\frac{\partial  E_\mathrm{m}}{\partial A}\right)_V \left\{\frac{\mathrm{d} A}{\mathrm{d} t}\right\}_{T,\ p,\ \kappa} ,
	\label{eq:dE_m_dt}
\end{split}
\end{equation}

\noindent which implies

\begin{equation}
	T\sigma_{\mathrm{tot}} =   \left(\Delta p\right)_\mathrm{eff} \left\{\frac{\mathrm{d} V}{\mathrm{d} t}\right\}_{T,\ p,\ \kappa} 
	+ \gamma_\mathrm{eff}\left\{\frac{\mathrm{d} A}{\mathrm{d} t}\right\}_{T,\ p,\ \kappa},
	\label{eq:sigma_tot_eff}
\end{equation}

\noindent where

\begin{equation}
	\left(\Delta p\right)_\mathrm{eff} = \Delta p - \left(\frac{\partial E_\mathrm{m}}{\partial V}\right)_A,
	\label{eq:Delta_p_eff}
\end{equation}

\noindent and

\begin{equation}
	\gamma_\mathrm{eff} = \gamma - \left(\frac{\partial E_\mathrm{m}}{\partial A}\right)_V.
	\label{eq:gamma_eff}
\end{equation}

\noindent Equation~(\ref{eq:sigma_tot_eff}) identifies the rate of entropy production for a growing vesicle as a sum of two pairs of forces and fluxes: the flow of volume
into the vesicle, coupled to a modified pressure difference across the membrane; and the rate of area increase of the membrane (due to accretion of lipids) which is coupled
to an modified surface tension.  Finally, the usual linear constitutive relation between the fluxes and forces may be invoked.  Making contact with \cite{VesRGMDFAJM} the
volume flux is written as

\begin{equation}
	\frac{1}{A}\left\{ \frac{\mathrm{d} V}{\mathrm{d} t} \right\}_{T,\ p,\ \kappa} = L_p \left( \Delta p \right)_\mathrm{eff} + L_\gamma \gamma_\mathrm{eff},
	\label{eq:constitutive}
\end{equation}

\noindent where $L_p$ and $L_\gamma$ are Onsager coefficients defined ``per unit area'' following the convention of the initial papers detailing the theory of hydraulic
conductivity \cite{KedemKatchalBiophys, KedemKatchalFaraday}.

In summary, we have shown how the growth of vesicles can be described thermodynamically, and reduced it under given conditions to the study of two-dimensional surfaces
whose volume changes according to Eq.~(\ref{eq:constitutive}). The rest of the paper is devoted to analyzing the growth such surfaces, and especially to their
deviations from a spherical shape.

\section{Membrane deformation} \label{sec:diff_geo}

In this Section, the formalism needed to describe surface deformations is outlined.  The aspects of differential geometry required are given in many standard
textbooks (e.g.~\cite{SchaumTenCalc}), but fortunately they are also used (in part) for the variational treatments which have been so prevalent in previous studies of
vesicle behavior.  The book by Ou-Yang \textit{et al.}~\cite{VesYangLiuXie} gives a good account of the details and the reader is referred to it for a discussion of the
mathematical background.  For this reason, \cite{VesYangLiuXie} forms the basis of the notation used below. 

\subsection{Surface geometry} \label{sec:diff_geo_background}

The two-dimensional surface (embedded in three-dimensions) which represents the membrane is defined by a vector field $\boldsymbol{r} = \boldsymbol{r} (u,v)$, where $u$ and
$v$ parametrize the surface. The tangent (vector) space associated with each point on the surface is then spanned by vectors $\boldsymbol{r}_l \equiv \partial
\boldsymbol{r} / \partial q_l$, where $l \in \{1, 2\},\:q_1 = u$ and $q_2 = v$.  From here, the first fundamental form, or metric, is defined as

\begin{equation}
	g_{lm} \equiv \boldsymbol{r}_l \cdot \boldsymbol{r}_m,
	\label{eq:metric}
\end{equation}

\noindent where the inverse metric $g^{ij}$ is defined such that $g^{ij}g_{jk} = \delta^i_k$, with $\delta^i_k$ the Kronecker delta symbol.  Here, $g$ is the determinant of
$g_{ij}$, given by

\begin{equation}
	g \equiv \frac{1}{2}\varepsilon^{lp}\varepsilon^{mq}g_{lm}g_{pq}, 
	\label{eq:g}
\end{equation}

\noindent where $\varepsilon^{ij}$ is an antisymmetric two-dimensional Levi-Civita symbol.  The determinant is used to define the surface area
element 

\begin{equation}
	\mathrm{d}A\equiv \sqrt{g}\mathrm{d}u\mathrm{d}v,
	\label{eq:dA}
\end{equation}

\noindent and the unit normal

\begin{equation}
	\hat{\boldsymbol{n}} \equiv \frac{\boldsymbol{r}_1 \times 
	\boldsymbol{r}_2}{\sqrt{g}}.
	\label{eq:normal}
\end{equation}

\noindent In order to quantify the curvature of a surface it is further necessary to define second derivatives $\boldsymbol{r}_{lm} \equiv \partial^2\boldsymbol{r} /
\partial q_{l}\partial q_{m}$, where the coefficients of the second fundamental form

\begin{equation}
	L_{lm} \equiv \boldsymbol{r}_{lm} \cdot \hat{\boldsymbol{n}},
	\label{eq:L_lm}
\end{equation}

\noindent allow us to make contact with (\ref{eq:mem_ener}) by writing

\begin{equation}
	H \equiv -\frac{1}{2}g^{ij}L_{ij}.
	\label{eq:H}
\end{equation}

\noindent For consistency with (\ref{eq:mem_ener}) and the majority of membrane related literature, (\ref{eq:H}) is defined here contrary to the usual convention of differential
geometry, so that the mean curvature of a sphere is positive, $H_{\mathrm{sphere}} = 1 / R$.

\begin{figure}[t]
\centering
\includegraphics[scale = 0.9]{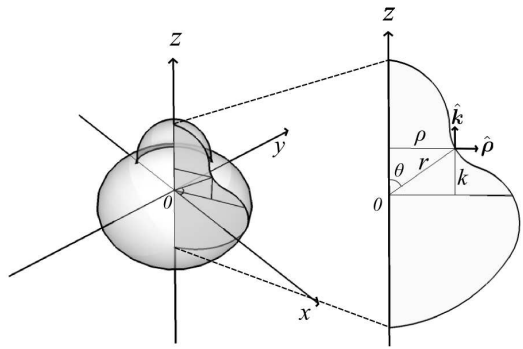}
\caption{Diagram showing the parametrization of axisymmetric shapes.}\label{fig:axisymmetric}
\end{figure}

With the basics in place we now focus on an explicit example of the above formalism: shapes which are invariant under rotation about the $z-$axis.  Such axisymmetric
vesicles have been the some of the most studied in variational calculations. The chosen parametrization is shown in Figure \ref{fig:axisymmetric}: here it is clear that
generic surface parameters $u$ and $v$ have been replaced by the familiar angles $\theta$ and $\phi$, where $0 \leq\theta\leq\pi$ is the inclination, and
$0\leq\phi<2\pi$ is the azimuthal angle.  In this case $\boldsymbol{r}$ has the form

\begin{equation}
	\boldsymbol{r} = \rho(\theta) \hat{\boldsymbol{\rho}} + k(\theta)\hat{\boldsymbol{k}},
	\label{axi_r}
\end{equation}

\noindent where $\hat{\boldsymbol{\rho}}=\cos{\phi}\hat{\boldsymbol{i}}+\sin{\phi}\hat{\boldsymbol{j}}$ and where $\hat{\boldsymbol{i}}, \hat{\boldsymbol{j}}$ and
$\hat{\boldsymbol{k}}$ are the unit vectors in the $x, y$ and $z$ directions respectively.  Tangent vectors are now given by

\begin{equation}
	\boldsymbol{r}_\theta = \rho'(\theta)\hat{\boldsymbol{\rho}} + k'(\theta)\hat{\boldsymbol{k}},\ \mathrm{and}\ \ \boldsymbol{r}_\phi = \rho(\theta)\hat{\boldsymbol{\phi}},
	\label{eq:r_theta,r_phi}
\end{equation}

\noindent where a dash is used as shorthand for the derivative with respect to $\theta$ and $\hat{\boldsymbol{\phi}} = -\sin{\phi}\hat{\boldsymbol{i}} +
\cos{\phi}\hat{\boldsymbol{j}}$. From the above it is clear the metric $g_{ij}$ is diagonal, and using (\ref{eq:metric}), (\ref{eq:g}), and (\ref{eq:normal}) it is a simple
exercise to show that

\begin{equation}
	\sqrt{g} = \rho\left[ \left( \rho' \right)^2 + \left( k' \right)^2 \right]^{1/2},
	\label{eq:rootg_axis}
\end{equation}

\noindent and

\begin{equation}
	\hat{\boldsymbol{n}} = \frac{\rho'\hat{\boldsymbol{k}} - k'\hat{\boldsymbol{\rho}}}{\left[ \left( \rho' \right)^2 + \left( k' \right)^2 \right]^{1/2}}.
	\label{eq:n_axis}
\end{equation}

\noindent Here, the explicit dependence on $\theta$ of functions $\rho$ and $k$ has been dropped for simplicity.  The form of second derivatives
$\boldsymbol{r}_{\theta\theta}$, $\boldsymbol{r}_{\theta\phi}$, $\boldsymbol{r}_{\phi\theta}$ and $\boldsymbol{r}_{\phi\phi}$ can also be calculated from which it is seen
that $L_{ij}$ is diagonal.  Inverting $g_{ij}$ it then follows from (\ref{eq:H}) that 

\begin{equation}
	H = \frac{1}{2}\left( \frac{k'\rho'' - k''\rho'}{\left[ \left( \rho' \right)^2 + \left( k' \right)^2 \right]^{3/2}} 
		- \frac{k'}{\rho\left[ \left( \rho' \right)^2 + \left( k' \right)^2 \right]^{1/2}} \right).
	\label{eq:H_axis}
\end{equation}

\subsection{Perturbation theory} \label{sec:pert}

The aim of this paper is to obtain stability conditions for deformations which take a vesicle from a sphere to an axisymmetric shape.  This can be achieved by writing the
shape dependent terms of constitutive relation (\ref{eq:constitutive}) as perturbations from a sphere and then comparing terms of equivalent order in the small parameter
controlling the perturbation.  The details of this are discussed in Section \ref{sec:stability}, however for now, in anticipation, perturbative expressions are required for
$A$, $V$, $\xi_1 \equiv \int H\mathrm{d}A$ and $\xi_2 \equiv \int H^2\mathrm{d}A$, \textit{i.e.}~the geometric terms which arise in (\ref{eq:constitutive}).  Choosing
appropriate forms for functions $\rho(\theta)$ and $k(\theta)$, we write

\begin{equation}
	\rho(\theta) = R\sin{\theta}\left( 1 + \epsilon(\theta) \right), 
	\label{eq:rho_perturb}
\end{equation}

\noindent and

\begin{equation}
	k(\theta) = R\cos{\theta}\left( 1 + \epsilon(\theta) \right),
	\label{eq:k_perturb}
\end{equation}

\noindent where $R$ is clearly the radius of the unperturbed sphere and perturbation $\epsilon(\theta)$---considered small on the scale of $R$---defines the resultant
axisymmetric shape.

Consider the surface area $A = 2\pi\int_0^\pi \sqrt{g}\mathrm{d}\theta$, this will serve as a template for calculating $V$, $\xi_1$ and $\xi_2$: the details of which are
confined to Appendix \ref{app:TenCalc}.  Substituting (\ref{eq:rho_perturb}) and (\ref{eq:k_perturb}) into (\ref{eq:dA}) and (\ref{eq:rootg_axis}) gives

\begin{equation}
	A = 2\pi R^2\int_0^\pi \mathrm{d}\theta \sin{\theta} \left[ 1 + 2\epsilon + \epsilon^2  
		+ \frac{1}{2}\left( \epsilon' \right)^2 + \mathcal{O}\left( \epsilon^4 \right)\right] ,
	\label{eq:A_1}
\end{equation}

\noindent where the explicit $\theta$ dependence of $\epsilon$ has been dropped and as before $\epsilon' = \mathrm{d} \epsilon / \mathrm{d} \theta$. Here, terms of order greater
than $\epsilon^3$ have been ignored, a choice which will become clear in the next section. Integrating the fourth term on the right-hand side gives  

\begin{equation}
	\int_0^\pi \mathrm{d}\theta \sin{\theta} \left( \epsilon' \right)^2 
		= -\int_0^\pi \mathrm{d}\theta\sin{\theta} \left[ \epsilon \hat{L}^2 \epsilon \right],
	\label{eq:int_(e')^2}
\end{equation}

\noindent where

\begin{equation}
	\hat{L}^2 \equiv \frac{1}{\sin{\theta}}\frac{\mathrm{d}}{\mathrm{d} \theta}\left( \sin{\theta}\frac{\mathrm{d}}{\mathrm{d} \theta} \right),
	\label{eq:L^2_theta}
\end{equation}

\noindent is the $\theta$-dependent part of the Laplacian in spherical polar coordinates.  With this in mind, (\ref{eq:A_1}) may now be computed by expanding $\epsilon$ in
terms of the zonal harmonics:

\begin{equation}
	\epsilon\left( \theta \right) = \varepsilon\sum_{l=2}^{\infty} a_l Y_l\left( \theta \right).
	\label{eq:sum_a_l_Y_l}
\end{equation}

\noindent Here, the zonal harmonics $Y_l\left( \theta \right)$ are the usual spherical harmonics with $m=0$.  The scale or size of the perturbation, $\varepsilon$, is
taken as a common factor of coefficients $\{a_l\}$.  For our purposes $a_0 = a_1 = 0$ as $Y_0 (\theta)$ and $Y_1(\theta)$ correspond to spherical growth and translation
respectively \cite{VesSaf}.  Using the properties of the zonal harmonics~\cite{Arfken}

\begin{equation}
	2\pi\int_0^\pi \mathrm{d}\theta\sin\theta\ Y_{l_1} (\theta) Y_{l_2} (\theta) = \delta_{l_1 l_2},
	\label{eq:zonal_orthogonal}
\end{equation}

\noindent and

\begin{equation}
	\hat{L}^2 Y_l (\theta) = -l(l+1) Y_l (\theta),
	\label{eq:L^2_Y_l}
\end{equation} 

\noindent it follows that

\begin{equation}
	A = 4\pi R^2 + \varepsilon^2 R^2\sum_{l=2}^\infty a_l^2 \left[ 1 + \frac{1}{2} l(l+1)\right] + \mathcal{O} (\varepsilon^4).
	\label{eq:A_2}
\end{equation}

\noindent As mentioned earlier, similar steps can be taken to write perturbative expressions for $\xi_1$, $\xi_2$ and $V$, however, in contrast to (\ref{eq:A_2}) these
results are cumbersome as they contain terms of third order in $\varepsilon$:  for example, the simplest result, V, is given by

\begin{equation}
\begin{split}
	V =& \frac{4}{3}\pi R^3 + \varepsilon^2 R^3 \sum_{l=2}^\infty a_l^2 \\
	& + \varepsilon^3\frac{R^3}{3}\sum_{l_1=2}^\infty \sum_{l_2=2}^\infty \sum_{l_3=2}^\infty a_{l_1} a_{l_2} a_{l_3} f(l_1,\ l_2,\ l_3)  \\
	&+ \mathcal{O}(\varepsilon^4),
	\label{eq:V_int_duplicate}
\end{split}
\end{equation}

\noindent where $f(l_1,\ l_2,\ l_3)$ is related to the square of a Wigner 3-j symbol.  The full details---including results for $\xi_1$ and $\xi_2$---can be found in
Appendix \ref{app:TenCalc}.

\section{Stability}\label{sec:stability}

We wish to address the question of vesicle stability \textit{dynamically}, and specifically, to determine when a
spherical vesicle becomes unstable and undergoes a shape change. Stability questions of this kind do not require the full dynamics of the system to be constructed and in our
case it is sufficient that only two variables are time-dependent: $R(t)$, the radius of the unperturbed sphere and $\varepsilon(t)$, which gives the magnitude of the
perturbation. The picture is the following. We study the growth of a vesicle which is spherical, but deformed by a small amount defined by (\ref{eq:sum_a_l_Y_l}). The
geometry of the perturbation---specified by the set $\{ a_{l} \}$---is fixed (time-independent), but the size of the perturbation---specified by $\varepsilon (t)$---is not. We
ask if there is a time (and so an $R(t)$) at which $\varepsilon (t)$ starts to increase with $t$, which will signal an instability.

Before carrying out this analysis, the growth mechanism needs to be specified, that is, the rate at which lipids are added to the membrane needs to be quantified. The
simplest assumption is that lipids attach themselves uniformly over the surface at a constant rate $\lambda$, such that

\begin{equation}
	\frac{\mathrm{d} A}{\mathrm{d} t}=\lambda A\ \Longrightarrow\ \ A(t) = A(0)e^{\lambda t}.
	\label{eq:growth}
\end{equation}

\noindent However, as described in \cite{VesRGMDFAJM}, when using a growth law of the above form it cannot be assumed that $R(t)$ is independent of $\varepsilon(t)$ so a
more consistent approach must be taken by moving to a new variable $r(t)$, defined as the radius of a sphere with equivalent surface area.  Setting $A=4\pi r^2$ and
comparing to (\ref{eq:A_2}) gives

\begin{equation}
	R = r\left(1-\varepsilon^2\frac{1}{8\pi}\sum_{l=2}^\infty a_l^2\left[1+\frac{1}{2}l(l+1)\right] + \mathcal{O} (\varepsilon^4)\right).
	\label{eq:R(r)}
\end{equation}

\noindent This can be substituted into the previous expressions for $\xi_1$ and $V$---(\ref{eq:xi_1_int}) and (\ref{eq:V_int_duplicate}) respectively---to give the results
(\ref{eq:xi_1_int_r}) and (\ref{eq:V_int_r}) while $\xi_2$ remains unchanged.   For clarity, we re-write geometric terms $\xi_1$, $\xi_2$ and $V$ in the following
simplified way

\begin{equation}
	\xi_1 = 4\pi r \left[ 1 + \varepsilon^2\xi_1^{(2)} + \varepsilon^3\xi_1^{(3)} + \mathcal{O} (\varepsilon^4) \right],
	\label{eq:xi_1_simple}
\end{equation}

\begin{equation}
	\xi_2 = 4\pi  \left[ 1 + \varepsilon^2\xi_2^{(2)} + \varepsilon^3\xi_2^{(3)} + \mathcal{O} (\varepsilon^4) \right],
	\label{eq:xi_2_simple}
\end{equation}

\noindent and

\begin{equation}
	V = \frac{4}{3}\pi r^3 \left[ 1 + \varepsilon^2 V^{(2)} + \varepsilon^3 V^{(3)} +\mathcal{O} (\varepsilon^4)\right],
	\label{eq:V_simple}
\end{equation}

\noindent where the time dependence is contained solely in variables $r (t)$ and $\varepsilon (t)$.  

At this stage we briefly note that writing $E_\mathrm{m} = 2\kappa\xi_2 - 2\kappa C_0 \xi_1 + \kappa C_0^2 A / 2$, and remembering $V=V(r, \varepsilon)$, implies that $E_m =
E_m (r, \varepsilon (r, V)) = E_m (A, V)$.  Therefore the term $\{\mathrm{d} E_\mathrm{m} / \mathrm{d} t \}_{T,\ p,\ \kappa}$ which arises in
(\ref{eq:sigma_tot_no_superscripts}) can indeed be written in the form (\ref{eq:dE_m_dt}) and so the constitutive relation (\ref{eq:constitutive}) is justified in the
context of a stability analysis.  

In order to calculate (\ref{eq:constitutive}) it is first necessary to write the partial derivatives $\left( \partial E_{\mathrm{m}} / \partial V \right)_{A}$ and $\left(
\partial E_{\mathrm{m}} / \partial A \right)_{V}$---which arise in the effective pressure and effective surface tension respectively---in terms of $r$ and $\varepsilon$.
As in \cite{VesRGMDFAJM} we have

\begin{equation}
	\left( \frac{\partial E_\mathrm{m}}{\partial V} \right)_{A} =
	\left( \frac{\partial E_\mathrm{m}}{\partial V} \right)_{r} =
	\left( \frac{\partial E_\mathrm{m}}{\partial\varepsilon} \right)_{r}  
	\left( \frac{\partial \varepsilon}{\partial V} \right)_{r}, 
	\label{eq:dEm_dV_A}
\end{equation}

\noindent and

\begin{equation}
\begin{split}
	\left( \frac{\partial E_\mathrm{m}}{\partial A} \right)_{V} &=
	\frac{1}{8\pi r}\,\left( \frac{\partial E_\mathrm{m}}{\partial r} \right)_{V} \\
	&= \frac{1}{8\pi r}\,\left\{ \left( \frac{\partial E_\mathrm{m}}
	{\partial r} \right)_{\varepsilon} + 
	\left( \frac{\partial E_\mathrm{m}}{\partial \varepsilon} \right)_{r}  
	\left( \frac{\partial \varepsilon}{\partial r} \right)_{V} \right\},
	\label{eq:dEm_dA_V}
\end{split}
\end{equation}

\noindent where it is worthwhile noting that since the volume $V$ has no terms of order $\varepsilon$, the membrane energy $E_\mathrm{m}$ must be taken to $\mathcal{O}
(\varepsilon^3)$ to ensure that partial derivative (\ref{eq:dEm_dV_A}) has the contributions of order $\varepsilon$ which are necessary to perform a stability analysis.
The calculation---to first order in $\varepsilon$---of partial derivatives $\left(\partial E_\mathrm{m} / \partial V\right)_A$ and $\left(\partial E_\mathrm{m} / \partial
A\right)_V$ is left to Appendix \ref{app:stability}: the results are given by (\ref{eq:dEm_dV_A_pert}) and (\ref{eq:dEm_dA_V_pert}) respectively.  In order to calculate the
left-hand side of (\ref{eq:constitutive}), the condition (\ref{eq:growth}) can be used to show that $\mathrm{d} r /\mathrm{d} t = \lambda r /2$, which, alongside
(\ref{eq:V_simple}) gives the result

\begin{equation}
	\frac{1}{A}\left\{ \frac{\textrm{d}V}{\textrm{d}t} \right\}_{T,\ p,\ \kappa} = \frac{\lambda r}{2} + \frac{2}{3} r V^{(2)} \varepsilon \left\{ \frac{\mathrm{d}
	\varepsilon}{\mathrm{d} t}\right\}_{T,\ p,\ \kappa} + \mathcal{O} (\varepsilon^2).
	\label{eq:LHSconstitutive}
\end{equation}

\noindent Substituting (\ref{eq:LHSconstitutive}) into (\ref{eq:constitutive}) and using the results of Appendix \ref{app:stability}, it is possible to find conditions on
the growth of the vesicle by equating terms of the same order in epsilon.

\subsection{Zeroth order: spherical growth}

At zeroth order in $\varepsilon$---equivalent to spherical growth---the condition which arises is given by

\begin{equation}
\begin{split}
	\frac{\lambda r}{2} =& L_p \left[ \Delta p - \frac{2\kappa}{r^3}  \left( 3\frac{\xi_2^{(2)}}{V^{(2)}} + C_0 r \right) \right] \\
	& + L_\gamma \left[ \gamma - \frac{\kappa}{2 r^2} \left(C_0 r\left( C_0 r - 4\right) - 6 \frac{\xi_2^{(2)}}{V^{(2)}}\right)\right].
	\label{eq:1/AdV_dt_zeroth}
\end{split}
\end{equation}

\noindent It can be seen from the definitions (\ref{eq:xi_2_simple}) and (\ref{eq:V_simple}) that the terms proportional to $\xi_2^{(2)} / V^{(2)}$ are dependent on the
choice of perturbation: something that should not be the case for an $\varepsilon = 0$ condition.  Therefore in order to make (\ref{eq:1/AdV_dt_zeroth}) independent of
perturbation it is necessary to impose the condition $L_\gamma = 2 L_p / r$.  This is the same identification which arose in \cite{VesRGMDFAJM} and leads to the same equation
for spherical growth in the linear regime: Eq.~(15) of \cite{VesRGMDFAJM}.  Indeed, if lipid accretion is ``turned off'' by setting $\lambda = 0$ then the equilibrium
condition for spherical vesicles is also recovered.  In our previous paper, the identification $L_\gamma = 2 L_p / r$  was motivated by knowing the spherical equilibrium
condition \textit{a priori}, here, the equilibrium result could have been derived independently by using the fact that the zeroth order condition must not rely on
perturbation choice by definition.   

\subsection{First order}

At first order in $\varepsilon$

\begin{equation}
\begin{split}
	\frac{\textrm{d}\varepsilon}{\textrm{d}t} =& \left( L_\gamma r - 2 L_p \right)\frac{27\kappa}{4 r^4 (V^{(2)})^2} \\
	&\times \left[ \left( \xi_2^{(3)} 
	- \frac{V^{(3)}\xi_2^{(2)}}{V^{(2)}} \right) - C_0 r\left( \xi_1^{(3)} + \frac{V^{(3)}}{3} \right)\right],
\end{split}
\label{eq:1/AdV_dt_first}
\end{equation}

\noindent which, after some manipulation, can be shown to be of the form

\begin{equation}
	\frac{\textrm{d}\varepsilon}{\textrm{d}t} = -\frac{L_\gamma \kappa C_0}{4r^4 \left(\xi_1^{(2)}\right)^2} \left(3\xi_1^{(3)} + V^{(3)}\right) (r-r_{c_1}) (r-r_{c_2}).
	\label{eq:1/AdV_dt_first_critical}
\end{equation}

\noindent Here, $r_{c_1} = 2L_p / L_\gamma$ and 

\begin{equation}
	r_{c_2} = \frac{3\xi_2^{(3)} \xi_1^{(2)} + V^{(3)} \xi_2^{(2)}}{C_0 \xi_1^{(2)} \left(3\xi_1^{(3)} + V^{(3)}\right)}
	\label{eq:r_c^2}
\end{equation}

\noindent are critical values of $r (t)$---the radius of a sphere with equivalent area---and correspond to critical values for the surface area $4\pi r_{c_1}^2$
and $4\pi r_{c_2}^2$ respectively.  Whether the surface area of the vesicle is greater than, less than or in between the two critical values of the surface area,
controls the sign of the right-hand side of (\ref{eq:1/AdV_dt_first_critical}) and therefore whether a particular perturbation is stable or unstable. 

\subsubsection{Single mode perturbations}

In order to understand the implications of (\ref{eq:1/AdV_dt_first_critical}) we consider a simplified case: perturbations which correspond to only one mode of the zonal
harmonics, that is $\epsilon (\theta) = \varepsilon a_l Y_l (\theta)$.  Substituting into previous results, Eqs.~(\ref{eq:xi_1_int_r}), (\ref{eq:V_int_r}) and
(\ref{eq:xi_2_int}) are reduced to

\begin{equation}
\begin{split}
	\xi_1 = 4\pi r &\left[1 - \frac{\varepsilon^2 a_l^2}{16\pi} \left[2-l(l+1)\right]\right. \\
	&\ \ -\frac{\varepsilon^3 a_l^3}{32\pi} f(l)\left[l(l+1)\left(2-l(l+1)\right)\right] \\
	&\ \  + \mathcal{O}(\varepsilon^4 )\Big],
	\label{eq:xi_1_single_a}
\end{split}
\end{equation}

\begin{equation}
\begin{split}
	\xi_2 = 4\pi & \left[ 1 -\frac{\varepsilon^2 a_l^2}{16\pi} \left[l(l+1)\left(2-l(l+1)\right)\right] \right. \\
	&\ \ +\frac{ \varepsilon^3 a_l^3}{16\pi} f(l)\left[l(l+1)\left(2 - l(l+1)\right)\right] \\
	&\ \ + \mathcal{O}(\varepsilon^4 )\Big],
	\label{eq:xi_2_single_a}
\end{split}
\end{equation}

\noindent and

\begin{equation}
\begin{split}
	V = \frac{4}{3}\pi r^3&\left[ 1 + \frac{3\varepsilon^2 a_l^2}{16\pi} \left[2-l(l+1)\right]\right. \\
	&\ \ + \frac{\varepsilon^3 a_l^3}{4\pi} f(l) + \mathcal{O} (\varepsilon^4 ) \Big],
	\label{eq:V_single_a}
\end{split}
\end{equation}

\noindent respectively, where $f(l) \equiv f(l,\ l,\ l)$.  Using (\ref{eq:xi_1_simple})-(\ref{eq:V_simple}) and substituting into (\ref{eq:1/AdV_dt_first_critical}) we write

\begin{equation}
	a_l\frac{\textrm{d}\varepsilon}{\textrm{d}t} = -\left( r - r_{c_1} \right) \left( r - r_{c_2} \right)  \frac{2\pi L_\gamma C_0 \kappa}{r^4} g(l),
	\label{eq:1/AdV_dt_first_single}
\end{equation}

\noindent where

\begin{equation}	
	r_{c_2} = \frac{20 l(l+1) - 6l^2(l+1)^2}{C_0\left(3l^2(l+1)^2 - 6l(l+1) + 8\right)},
	\label{eq:r_c_2_single}
\end{equation}

\noindent and

\begin{equation}
	g(l) = \frac{f(l) \left(3l^2(l+1)^2 - 6l(l+1) + 8\right)}{\left( 2-l(l+1) \right)^2}.
	\label{eq:g(l)}
\end{equation}

\noindent It is helpful at this stage to introduce dimensionless quantities: radii are re-scaled by a factor of $C_0$ such that $\tilde{r} = C_0 r$, $\tilde{r}_{c_1} = C_0
r_{c_1}$ and $\tilde{r}_{c_2} = C_0 r_{c_2}$ whilst time is re-scaled to give $\tau = \lambda t$.  Inserting this into the stability condition gives rise to a natural choice for re-scaling the Onsager coefficient associated with surface
growth, $\tilde{L}_\gamma = L_\gamma \kappa C_0^3 / \lambda$.  Also, noting that $g(l)$ is always positive and $\tilde{r}_{c_2} (l)$ is always negative for $l\geq 2$
it can be seen that for such perturbations---corresponding to a single mode of the zonal harmonics--- there is only one critical surface area, $4\pi r_{c_1}^2$.  Taking
this into account (\ref{eq:1/AdV_dt_first_single}) becomes

\begin{equation}
	a_l\frac{\textrm{d}\varepsilon}{\textrm{d}\tau} = -\left( \tilde{r} - \tilde{r}_{c_1} \right) \left( \tilde{r} + \vert\tilde{r}_{c_2}\vert \right)  \frac{2\pi
	\tilde{L}_\gamma}{\tilde{r}^4} g(l).
	\label{eq:depsilon_dt_rescaled}
\end{equation}

\noindent In addition, it should be noted that since $f(l,\ l,\ l)$ is zero for odd values of $l$, perturbations which correspond to odd $l$ are constant in time for all
radii at first order.  In order to investigate the stability of odd zonal harmonics it is necessary to take the analysis presented to next order in $\varepsilon$ which is
left for future work.  Indeed, at a heuristic level, it might have been anticipated that the stability of harmonic perturbations which are asymmetric about $\theta=0$ are
determined at an order greater than symmetric perturbations, as a better ``resolution'' is needed to differentiate between shapes with lower symmetry.

In order to interpret the single-mode stability condition (\ref{eq:depsilon_dt_rescaled}) further, it is first necessary to assume some typical values of the constants
involved: we estimate a value for $\kappa$ of $10^{-19}$J \cite{Marsh2006} and a value for $C_0$ of $10^{7}\mathrm{m}^{-1}$ such that $C_0 r$ is of order 1 for a 100nm
spherical vesicle. Following \cite{FanelliMckaneReplyComment} we use a value for $L_p$, the hydraulic permeability, of $7.5 \times
10^{-13}\mathrm{m}\mathrm{s}^{-1}\mathrm{Pa}^{-1}$.  What is \textit{not} known is a typical value for $L_\gamma$, the Onsager coefficient linking surface tension to the
rate of change in the volume of the interior. 

It is possible to deduce an estimate for $L_\gamma$ by recalling Eq.~(\ref{eq:constitutive}) and comparing the relative contributions that both pressure and surface tension
terms make to the rate of change of volume.  In order to do this it is necessary to estimate an order of magnitude for the effective pressure difference $(\Delta
p)_{\mathrm{eff}}$ and effective surface tension $\gamma_{\mathrm{eff}}$.  For simplicity we ignore the modifications due to the membrane energy and drop the subscript
\textit{effective}. An estimate of $\gamma = 10^{-3}\mathrm{N}\mathrm{m}^{-1}$ is provided by \cite{LateralTensionGruen} though estimating the pressure difference is less
clear.  Following \cite{FanelliMckaneReplyComment} we ask: what is the pressure difference which maintains mechanical equilibrium in a growing vesicle?  Using Eq.~(15) of
\cite{VesRGMDFAJM}, 

\begin{equation}
	\Delta p = \frac{\lambda r}{2 L_p} + \frac{C_0 \kappa }{r^2}\left( C_0 r -2 \right) - \frac{2\gamma}{r},
	\label{eq:delta_p_mechanical}
\end{equation}

\noindent it is possible to use an estimate for $\lambda$ along with those previously taken for $C_0$, $\kappa$, $\gamma$ and $L_p$ to compute the pressure difference
needed to maintain equilibrium in a vesicle with $r = 100\mathrm{nm}$.  Taking $\lambda = 10^{-4}\mathrm{s}^{-1}$ (the middle of the range proposed in
\cite{SvetinaBozicBiophys}) gives an estimate for the pressure difference of 0.1-0.01 bar.  Going back to (\ref{eq:constitutive}), we argue that for such processes the term
$L_\gamma \gamma$ will be neither negligible nor significantly larger than the term $L_p \Delta p$. Indeed, for the purposes of an estimate we require that the two terms
are the same order.  Taking a pressure difference of 0.1 bar this assumption implies that $L_\gamma$ is of the order $10^{-6}\mathrm{s}^{-1} \mathrm{Pa}^{-1}$.
As a check, we may calculate an order of magnitude for the re-scaled quantity $\tilde{L}_\gamma$ using the values above.  This leads to an order of magnitude for $\tilde{L}_\gamma$
of one.  For completeness, using these estimates the dimensionless analogue for the hydraulic conductivity, $\tilde{L}_p= L_p \kappa C_0^4 / \lambda$  is therefore 7.5.

\begin{figure}[t] 
	\centering 
	\includegraphics[scale = 1]{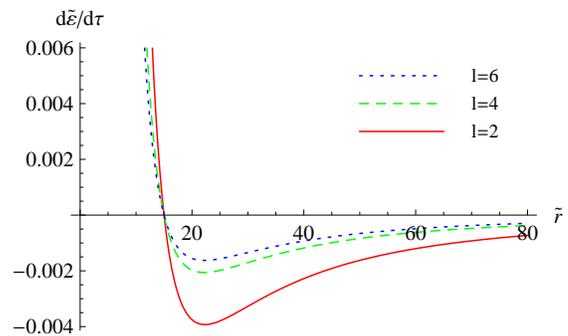} 
	
	\caption{(Color online) Graph showing the growth of three separate perturbations to a spherical vesicle at different radii.  Three specific perturbations are shown: the
	zonal harmonics corresponding to $l=2$ (solid), $l=4$ (dashed) and $l=6$ (dotted).  As highlighted in the text, order of magnitude estimates for physical constants have
	been used such that the dimensionless quantities $\tilde{L}_\gamma$ and $\tilde{L}_p$ are taken to be 1 and 7.5 respectively.  In addition, coefficients $a_l$ are taken
	to be positive.}
	
	\label{fig:single_mode_rescale} 
\end{figure}

Using this estimate for $\tilde{L}_\gamma$ it is possible to plot curves which show the stability of particular perturbations at different radii.
Fig.~\ref{fig:single_mode_rescale} plots $\mathrm{d}\tilde{\varepsilon} / \mathrm{d} \tau$ against $\tilde{r}$, where $\tilde{\varepsilon} = a_l \varepsilon$ ($a_l >0$),
for three specific perturbations: the zonal harmonics corresponding to $l=2$, $l=4$ and $l=6$.  First, we note that all curves cross the axis at $\tilde{r}_{c_1} = 2C_0 L_p
/ L_\gamma$, which in the framework of estimates discussed above is 15.  Below this point, all modes are unstable---that is, $\mathrm{d}\tilde{\varepsilon} / \mathrm{d}
\tau$ is positive---with lowest values of $l$ being the most unstable \textit{i.e.}~growing at the fastest rate.  Above the critical point all modes are stable (decay in
time) with lowest values of $l$ decaying fastest.  However, we may note that the inverse is true for perturbations defined opposite to those discussed, that is $a_l
\longrightarrow -a_l$.  These perturbations are stable beneath $\tilde{r}_{c_1}$ (with low $l$ modes decaying fastest) and unstable above $\tilde{r}_{c_1}$ (with low $l$
modes growing fastest).  

We note that the curves shown in Fig.~\ref{fig:single_mode_rescale} are asymmetric about $\tilde{r}_{c_1}$.  That is, the scale of perturbation growth beneath the critical
radius is much larger than the scale of decay above the critical radius.  Similarly, for $a_l<0$, decay beneath $\tilde{r}_{c_1}$ is much larger than growth above
$\tilde{r}_{c_1}$.  However, as we will discuss in Section \ref{sec:conclusions}, before a more comprehensive analysis may be carried out, the estimates of the physical
parameters used need to be significantly improved.  As such, a more detailed discussion of the implications of this feature is left for future work.

\subsubsection{Ellipsoidal perturbations}

We may ask whether (\ref{eq:depsilon_dt_rescaled}) can be reduced to our previous results for (axisymmetric) ellipsoidal deformations \cite{VesRGMDFAJM}.  In terms of the
dimensionless quantities introduced above, the stability condition in that paper is given by

\begin{equation}
	\frac{\mathrm{d} \varepsilon}{\mathrm{d} \tau} = -\left( \tilde{r} - \tilde{r}_{c_1} 
	\right) \left( 5\tilde{r} + 6 \right)  \frac{2\pi 
	\tilde{L}_\gamma}{\tilde{r}^4}\frac{15}{28\pi\left( c_1 - a_1 \right)},
	\label{eq:depsilon_dt_ellipsoid}
\end{equation}

\noindent where the parametrization used to characterize (axisymmetric) ellipsoidal perturbations may be written as

\begin{equation}
	\begin{split}
		\rho_\mathrm{el} &= R\left(1+a_1 \varepsilon\right)\sin{\theta},\\ k_\mathrm{el} 
		&= R\left(1+c_1 \varepsilon\right)\cos{\theta}.
	\label{eq:rho_k_ellipsoid}
\end{split}
\end{equation}

\noindent Here, the notation of Section \ref{sec:pert} and Fig. \ref{fig:axisymmetric} has been used: the points on the surface which are an angle $\theta$ from the
positive $z$-axis are a distance $\rho_\mathrm{el}$ from the $z$-axis and $k_\mathrm{el}$ from the $x$-$y$ plane.   Such a parametrization is different to the more general
approach taken in the main part of this paper, however, it is possible to compare the two at first order in $\varepsilon$.  In order to do so, it is necessary to move to
variable $r$, the radius of a sphere of equivalent area.  Using Eq.~(17) of \cite{VesRGMDFAJM} we see that, to first order in $\varepsilon$,  the surface area of an
ellipsoid with parametrization (\ref{eq:rho_k_ellipsoid}) is given by

\begin{equation}
	A = 4\pi R^2 \left[ 1 + \frac{2}{3}\left( 2a_1 + c_1 \right)\varepsilon + \mathcal{O} 
	(\varepsilon^2)\right].
	\label{eq:A_ellipsoid}
\end{equation}

\noindent Setting $A = 4\pi r^2$ gives

\begin{equation}
	R = r\left[1 - \frac{1}{3}\left( 2a_1 + c_1 \right)\varepsilon + \mathcal{O} 
	(\varepsilon^2)  \right],
	\label{eq:R(r)_ellipse}
\end{equation}

\noindent which can be substituted back into (\ref{eq:rho_k_ellipsoid}) to give

\begin{equation}
	\begin{split}
		\rho_\mathrm{el} &= r\left[1+ \frac{1}{3} \left(a_1 - c_1\right)\varepsilon + 
		\mathcal{O} (\varepsilon^2)\right]\sin{\theta},\\ k_\mathrm{el} &= r\left[1 - 
		\frac{2}{3} \left(a_1 - c_1\right)\varepsilon + \mathcal{O} 
		(\varepsilon^2)\right]\cos{\theta}.
		\label{eq:rho_k_ellipsoid_r}
	\end{split}
\end{equation}

\noindent Here, the radial distance to a point on the surface of the deformed shape (ellipsoid) is given by

\begin{equation}
	\sqrt{\left( \rho_\mathrm{el} \right)^2 + \left( k_\mathrm{el} \right)^2} = r\left[ 1 
	+ \frac{1}{3}\left( c_1-a_1 \right)\varepsilon\left( 3\cos^2 \theta - 1 \right) + 
	\mathcal{O} (\varepsilon^2) \right],
	\label{eq:radius_ellipsoid}
\end{equation}

\noindent where taking $Y_2 (\theta) = \frac{1}{4}\sqrt{\frac{5}{\pi}} \left( 3\cos^2 \theta -1 \right)$~\cite{Arfken} this may be re-written as

\begin{equation}
	\sqrt{\left( \rho_\mathrm{el} \right)^2 + \left( k_\mathrm{el} \right)^2} = r\left[ 1 
	+ \frac{4}{3}\sqrt{\frac{\pi}{5}}\left( c_1-a_1 \right)\varepsilon Y_2 (\theta) + 
	\mathcal{O} (\varepsilon^2) \right].
	\label{eq:radius_ellipsoid_Y}
\end{equation}

\noindent With this in place we may turn to the general axisymmetric parametrization set out earlier in this paper.  Noting from (\ref{eq:R(r)}) that $R = r\left[ 1 +
\mathcal{O} (\varepsilon^2) \right]$ Eqs.~(\ref{eq:rho_perturb}) and (\ref{eq:k_perturb}) may be written

\begin{equation}
	\begin{split}
		\rho &= r\left[1 + \epsilon (\theta) + \mathcal{O} (\varepsilon^2) 
		\right]\sin{\theta},\\ k &= r\left[1 + \epsilon (\theta) + \mathcal{O} 
		(\varepsilon^2)\right]\cos{\theta},
		\label{eq:rho_k_r}
	\end{split}
\end{equation}

\noindent where the reader is reminded that the function $\epsilon (\theta)$ is of order $\varepsilon$.  From here it follows from the calculation of $\sqrt{\rho^2 + k^2}$
that the two parameterizations are equivalent to first order in $\varepsilon$ if 

\begin{equation}
	\epsilon (\theta) = \frac{4}{3}\sqrt{\frac{\pi}{5}}\left( c_1 - a_1 \right)\varepsilon 
	Y_2 (\theta),
	\label{eq:epsilon(theta)_ellipsoid}
\end{equation}

\noindent which implies

\begin{equation}
	a_2 = \frac{4}{3}\sqrt{\frac{\pi}{5}}\left( c_1 - a_1 \right).
	\label{eq:a_2}
\end{equation}

\noindent Substituting this expression for $a_2$ into (\ref{eq:depsilon_dt_rescaled}) and using the facts that $g(2)= (5/7)\sqrt{5/\pi}$ and $\tilde{r}_{c_2}(l=2)= -5/6$,
the stability condition (\ref{eq:depsilon_dt_ellipsoid}) for ellipsoidal deformations can be recovered.

\section{Conclusions and discussion}\label{sec:conclusions}

The purpose of this paper has been two-fold. Firstly, to systematically set up the 
thermodynamic description of vesicle growth, and secondly, to analyze the stability of 
deformations: specifically, to determine when spherical surfaces are unstable to small 
axisymmetric perturbations.

While thermodynamic descriptions of vesicle growth based on LNET have been discussed 
previously, the form of the fluxes and forces were obtained through physical arguments, 
and not derived from a comprehensive analysis of the thermodynamics of a membrane bilayer 
in an aqueous environment.  The analysis presented in Section \ref{sec:thermodynamics} and 
Appendix \ref{app:discontinuous} aims to do just this: it is based on the concept of a 
discontinuous system discussed by de Groot and Mazur \cite{NonEquilThermDeGrootMazur}, 
although considerably elaborated for a vesicle system.  It is assumed that as lipids are 
incorporated into the membrane, the entropy produced---due to the realignments of 
others---can be neglected.  However, the energy change that accompanies such accretion 
cannot be ignored.  Indeed, the assumption of the Canham-Helfrich-Evans 
form---Eq.~(\ref{eq:mem_ener})---replaces the complexity of the lipid bilayer by an energy 
defined in terms of the geometrical properties of a single surface.

In order to investigate the implications of this thermodynamic approach, we focused on the 
stability of vesicles growing due to accretion.  In the context of such a stability 
analysis, the energy is assumed to rely on two time-dependent variables: surface area 
$A(t)$ and volume enclosed $V(t)$.  Here, in analogy to a standard two-component system 
partitioned into two regions, the entropy produced is characterized by an effective 
pressure difference and effective surface tension.  The effective pressure is just the 
normal pressure difference modified by the term $\left( \partial E_\mathrm{m} /\partial V 
\right)_A$, and the effective surface tension is the normal surface tension modified by 
the term $\left( \partial E_\mathrm{m} /\partial A \right)_V$.

Whilst undoubtedly such a thermodynamic approach can be improved upon, the nature of the 
approximations made are argued for on physical grounds, and specific physical 
justifications have been given where possible. At the very least, the nature and extent of 
the assumptions are clearly visible providing a starting point for attempts to relax them. 

With the study reduced to the dynamics of a surface of area $A(t)$ enclosing a volume 
$V(t)$ and characterized by an energy $E_\mathrm{m} (t)$---given by 
(\ref{eq:mem_ener})---the problem is geometrical in nature and differential geometry may 
be used.  In general the approach is applicable to any smooth surface with no particular 
symmetry properties, but the analysis that we present in Section \ref{sec:diff_geo} is 
restricted to an axisymmetric surface. The reasons for making this choice are two-fold: it 
is common in variational studies of vesicle growth, and, for reasons of simplicity.  
However there is no problem in principle in treating a general shape. The main difference 
is that the perturbation would depend on the angle $\phi$ as well as $\theta$, and would 
be spanned by the full set of spherical harmonics $Y_{l,m}(\theta,\phi)$. This would lead 
to sums on $l$ (for instance in Eqs.~(\ref{eq:A_2}) and (\ref{eq:V_int_duplicate}))
being replaced by sums on both $l$ and $m$.

When carrying out the stability analysis in Section \ref{sec:stability} we further 
specialized to perturbations about a spherical vesicle, but once again other unperturbed 
geometries could be chosen. A peculiarity of the perturbation expansion which has already 
been remarked on in \cite{VesRGMDFAJM} is that it is necessary to develop the perturbation 
expansion to third order in order to find the growth rate of $\varepsilon(t)$ to leading 
order. This unfortunately makes the calculation more complicated than might have naively 
been expected.  Nevertheless, the present analysis places no constraint on the nature of 
the perturbation, other than it is axisymmetric. This is in contrast to our earlier 
treatment \cite{VesRGMDFAJM}, which assumed that the distorted surface was an ellipsoid.  
The results of the present treatment are shown to reduce to those of \cite{VesRGMDFAJM} 
for the case where the perturbation corresponds to the $l=2$ zonal harmonic.

In general it is found that there are two critical radii at which changes in stability 
occur.  However, in the case when perturbations take the form of a single zonal harmonic, 
only one of these radii is physical: that given by $2L_p / L_\gamma$, where $L_p$ is the 
hydraulic conductivity and $L_\gamma$ the Onsager coefficient corresponding to changes in 
surface area due to lipid accretion.  In order to quantify our results we made an estimate 
for $L_\gamma$ using simple physical arguments.  For a more complete approach this 
phenomenological coefficient would need to be measured experimentally.  Using order of 
magnitude estimates we see that---under the conditions of accretion proportional to 
surface area--- spherical vesicles at radii other than the critical radius are always 
unstable.  Perturbations corresponding to zonal harmonics of lowest $l$ grow at the 
fastest rate (are the most unstable) with modes corresponding to larger $l$ becoming 
increasingly stable.

There are many ways in which the analysis presented here could be taken forward. The most 
pressing need is for further experimental studies against which the predictions made in 
this paper can be compared.  Experimental work regarding vesicle shape changes has 
previously been carried out though the focus has been on transitions induced by either 
temperature \cite{JKES91} or osmotic pressure changes \cite{Pen01}. A summary of some 
relevant experimental studies was given in \cite{VesRGMDFAJM}; of particular interest is 
\cite{Pen01}, where initially symmetric vesicles were found to deform into oblate shapes 
and then prolate ones under osmotic pressure changes which serve to reduce the enclosed 
volume.  

In relation to these existing experimental studies, it is appropriate to scrutinize our 
choice of driving mechanism.  Here, we assumed that lipids accrete to the surface at a 
constant rate per unit area.  Although this leads to a growth law of exponential form 
(\ref{eq:growth}), it is worth highlighting that $\lambda$ is taken to be very small (of 
the order $10^{-4}\mathrm{s}^{-1}$~\cite{SvetinaBozicBiophys}).  That is, for the range of 
observed vesicle sizes, surface growth will still be very slow.  Our choice of mechanism 
was influenced by a number of factors.  Firstly, experiments of this type have been 
carried out and are documented in~\cite{LifeLuisi}.  Indeed, the suggestion is that growth 
due to accretion is a plausible phenomenon in the pre-biotic scenario in which we are 
interested.  Secondly, this approach already entails a great deal of mathematics; we 
therefore wanted to implement the simplest choice for driving the system away from 
equilibrium.  However, growth laws of a more complicated form could be incorporated if 
necessary.  Finally, we were also conscious to demonstrate that both the results presented 
here and the extended theoretical background detailed in Section~\ref{sec:thermodynamics} 
are connected to our previous paper~\cite{VesRGMDFAJM}.   Furthermore, in reality, lipids 
in solution are likely to form micelles and small vesicles.  However, the additional 
effects of such micelle-vesicle or vesicle-vesicle adsorption are not considered here.  
With these points in mind, a more experimentally accessible approach might be to extend 
previous work so that temperature change---rather than the accretion of lipids---is the 
effect which gives rise to shape changes.  However, the analysis needed to incorporate 
such a feature presents a further technical challenge and, in light of the already lengthy 
theoretical background, it is left for future work.

In addition to the above, it is clear from experiment that non-axisymmetric vesicles are 
observed (see for example~\cite{KBJK+90}) and so our analysis should be extended to 
investigate transitions from a sphere to an arbitrary shape, as opposed to those invariant 
about the $z$-axis.  Similarly, a more general theory based on deformations from an 
arbitrary shape (as opposed to a sphere) could also be developed.  Lastly, more realistic 
models of the membrane such as the Area-Difference Elasticity model \cite{miao94} could be 
used.

All these studies would benefit from more experimental input. However, the foundations we 
have laid, and the methods we have developed in this paper, do allow for these more 
general analyzes to be carried out. We expect that they will lead to a more comprehensive 
understanding of the dynamics of vesicle growth in the near future.

\begin{acknowledgments}
We wish to thank Duccio Fanelli for useful discussions. RGM wishes to thank the EPSRC (UK) for support.
\end{acknowledgments}

\appendix

\section{Entropy production}\label{app:discontinuous}

In this Appendix, the calculation of entropy production is summarized.  For further background the reader is referred to chapter 15 of \cite{NonEquilThermDeGrootMazur} from
which this analysis is adapted.  As usual, subscript $k \in \{w,\ l\}$ denotes the component (water and lipids respectively) and superscript $\alpha \in \{\mathrm{I} -
\mathrm{V}\}$ denotes the region (see Figure \ref{fig:schematic}).  Starting with the total entropy in each region $\alpha$

\begin{equation}
	S^\alpha \equiv  \int_\alpha \rho s \mathrm{d} V,
	\label{eq:S^alpha}
\end{equation}

\noindent the rate-of-change of $S^\alpha$ can be written as a sum of entropy fluxes over the boundary and any entropy produced in the bulk  

\begin{equation}
\begin{split}
	\frac{\mathrm{d} S^\alpha}{\mathrm{d} t} = \int_\alpha \frac{1}{T} &
		\Big( -\vect{J}_{\mathrm{q}} - \vect{J}_{\mathrm{nem}} + \sum_k \mu_k \rho_k 
		\left( \vect{v}_k - \vect{v}^\mathrm{b} \right) \\
	& \ \ - \rho h \left( \vect{v} - 
	\vect{v}^\mathrm{b}\right)\Big)\cdot\mathrm{d}\vect{A} + \int_\alpha\sigma\mathrm{d} 
	V.
	\label{eq:dS^alpha_dt_2}
\end{split}
\end{equation}

\noindent Here, entropy fluxes have been taken in the traditional hydrodynamic form (Equation (20), chapter 3 of \cite{NonEquilThermDeGrootMazur}) plus a term
$\vect{J}_{\mathrm{nem}}/T$ which arises due the nematic nature of the lipid molecules~\cite{ConservationLiquidCrystalsEricksen}.  The other symbols have their usual
meanings~\cite{NonEquilThermDeGrootMazur}, $\boldsymbol{J}_{\mathrm{q}}$ is heat flow, $h = u + p\nu$ is the enthalpy, $\boldsymbol{v}^\mathrm{b}$ is the velocity of the
boundary---zero for all impermeable boundaries---and $\boldsymbol{v} = \sum_k \boldsymbol{v}_k \rho_k$ defines both total and partial velocities.  At this stage, it is
assumed that entropy fluxes associated with the nematic nature of the lipids are tangent to the boundary at any point, and so make no contribution to the rate of change of
entropy:
		
\begin{equation}	
	\int_\alpha \vect{J}_{\mathrm{nem}}\cdot \mathrm{d} \vect{A} = 0.
	\label{eq:int_J_nem_dA=0}
\end{equation}

\noindent In the same manner as set out in chapter 15 of~\cite{NonEquilThermDeGrootMazur}---though adapted for the system considered in Fig.
\ref{fig:schematic}---Eq.~(\ref{eq:int_J_nem_dA=0}) can be used to write an expression for the total rate of entropy produced in the system 
	
\begin{equation}
\begin{split}
	\sigma_{\mathrm{tot}} 
	= \frac{1}{T}\sum_{\alpha = \mathrm{I}}^{\mathrm{III}} \int_\alpha &\bigg( -\vect{J}_{\mathrm{q}}  + \sum_k \mu_k \rho_k \left(\vect{v}_k -
	\boldsymbol{v}^\mathrm{b}\right) \\ 
	&\ \  - \rho h \left( \boldsymbol{v} - \boldsymbol{v}^\mathrm{b}\right) \bigg)\cdot\mathrm{d}\boldsymbol{A},
	\label{eq:sigma_tot}
\end{split}
\end{equation}

\noindent where integration is now over internal (permeable) boundaries only.  Again, adapted from \cite{NonEquilThermDeGrootMazur}, conservation of internal energy and
conservation of mass are given by

\begin{equation}
	\sum_{\alpha=\mathrm{I}}^{\mathrm{III}} \int_\alpha \left( \vect{J}_\mathrm{q} + \rho h\left(\boldsymbol{v} - \boldsymbol{v}^\mathrm{b}\right)\right)\cdot\mathrm{d}\vect{A} = 0,
	\label{eq:conserveinternalenergy}
\end{equation}

\noindent and

\begin{equation}
	\sum_{\alpha=\mathrm{I}}^{\mathrm{III}}\int_\alpha \rho_k \left(\boldsymbol{v}_k - \boldsymbol{v}^\mathrm{b}\right)\cdot\mathrm{d}\vect{A}
		= - \sum_{\alpha=\mathrm{I}}^{\mathrm{III}} \frac{\mathrm{d} M^\alpha_k}{\mathrm{d} t} = 0,
	\label{eq:conservemass}
\end{equation}

\noindent respectively.  Imposing (\ref{eq:conserveinternalenergy}) on (\ref{eq:sigma_tot}) and using the facts that mass fluxes are assumed
to be evenly distributed across boundaries and velocities are taken to be in the normal direction (see Section \ref{sec:thermodynamics}) gives 

\begin{equation}
	\sigma_{\mathrm{tot}} = -\frac{1}{T}\sum_{\alpha=\mathrm{I}}^{\mathrm{III}} \sum_k \frac{\mathrm{d} M^\alpha_k}{\mathrm{d} t} \bar{\mu}_k^\alpha,
	\label{eq:extra}
\end{equation}

\noindent where a bar above a variable is used to denote ``average over a boundary'' in the sense of (\ref{eq:bar_x}).  (Note that for uniform regions I and III,
$\bar{\mu}_k^\alpha = \mu_k^\alpha$).  Using (\ref{eq:conservemass}) to eliminate mass flows out of region I, the exterior, results in (\ref{eq:sigma_tot_main}) which is
re-written here using the difference notation (\ref{eq:Delta_alpha_beta})

\begin{equation}
	\sigma_{\mathrm{tot}} = \frac{1}{T}\sum_k \sum_{\alpha=\mathrm{II}}^{\mathrm{III}} 
	\left( \Delta_{\mathrm{I} ,\ \alpha\ }  \bar{\mu}_k \right)\frac{\mathrm{d}M^\alpha_k}{\mathrm{d} t}.
	\label{eq:sigma_tot_3}
\end{equation}

\noindent As outlined in Section \ref{sec:thermodynamics} averages over the boundary to the membrane are replaced by averages over the neutral surface.  Furthermore, it is
possible to expand chemical potential differences in terms of other thermodynamic variables.  Using (\ref{eq:Deltamu_k}) gives

\begin{equation}
\begin{split}
	\sigma_{\mathrm{tot}} = \frac{1}{T}\sum_k \sum_{\alpha=\mathrm{II}}^{\mathrm{III}} 
	& \Big( \bar{\nu}_k^\alpha \left( \Delta_{\mathrm{I} ,\ \alpha\ }p \right) 
	+ \bar{\psi}_k^\alpha \left( \Delta_{\mathrm{I} ,\ \alpha\ }\kappa \right) \\
	& \ \ + \{\Delta_{\mathrm{I} ,\ \alpha\ }\bar{\mu}_k\}_{T,\ p,\ \kappa}\Big)\frac{\mathrm{d}M^\alpha_k}{\mathrm{d} t}.
	\label{eq:sigma_tot_4}
\end{split}
\end{equation}

\noindent So the entropy produced has been written as a sum of thermodynamic forces---differences of variables across discontinuities---and thermodynamic fluxes.  We proceed
by considering each term separately.  The first term is summed over regions III and II, the interior and the membrane respectively.  Consider first the interior: a uniform
region, we may write $V^{\mathrm{III}} =  V^{\mathrm{III}}\left(T,\ p,\ \{M^\mathrm{III}_k\},\ \kappa\right)$ and therefore
	
\begin{equation}
	\begin{split}
		\left\{ \frac{\mathrm{d} V^\mathrm{III}}{\mathrm{d} t}\right\}_{T,\ p,\ \kappa} &= \sum_k \left( \frac{\partial V^\mathrm{III}}{\partial M^\mathrm{III}_k}\right)_{T,\ p,\
		\kappa,\ \{M^\mathrm{III}_{i\neq k}\}} \frac{\mathrm{d}M^\mathrm{III}_k}{\mathrm{d} t} \\
		&= \sum_k \nu^\mathrm{III}_k \frac{\mathrm{d}M^\mathrm{III}_k}{\mathrm{d} t}.
	\end{split}
\end{equation}

\noindent Consider now the term relating to the membrane (a non-uniform region): from (\ref{eq:j_k^alpha}) we have

\begin{equation}
	\sum_k \bar{\nu}_k^\mathrm{II} \frac{\mathrm{d} M_k^\mathrm{II}}{\mathrm{d} t} = - \sum_k\bar{\nu}_k^\mathrm{II}\int_\mathrm{II} \rho_k \left( \boldsymbol{v}_k -
	\boldsymbol{v}^\mathrm{b} \right)\cdot\mathrm{d}\boldsymbol{A}.
	\label{eq:nu^2_k_dM^2_k_dt}
\end{equation}

\noindent Recalling that mass fluxes per unit area are assumed constant, using the definition of $\bar{\nu}_k^\mathrm{II}$ and the fact that $\sum_k \rho_k \nu_k = 1$ (see
Appendix II of \cite{NonEquilThermDeGrootMazur})  gives

\begin{equation}
	\begin{split}
		\sum_k \bar{\nu}_k^\mathrm{II} \frac{\mathrm{d} M_k^\mathrm{II}}{\mathrm{d} t} &= -\sum_k \int_\mathrm{II} \nu_k \rho_k \left( \boldsymbol{v}_k
		-\boldsymbol{v}^\mathrm{b} \right)\cdot\mathrm{d}\boldsymbol{A} \\
		&= \int_\mathrm{II} \boldsymbol{v}^\mathrm{b}\cdot\mathrm{d}\boldsymbol{A} - \sum_k \int_\mathrm{II} \nu_k\rho_k\boldsymbol{v}_k\cdot\mathrm{d}\boldsymbol{A}\\
		&=\frac{\mathrm{d} V^\mathrm{II}}{\mathrm{d} t} - \sum_k \int_\mathrm{II} \nu_k\rho_k\boldsymbol{v}_k\cdot\mathrm{d}\boldsymbol{A}.
	\end{split}
	\label{eq:nu^2_k_dM^2_k_dt_2}
\end{equation}

\noindent Here, we recognize that the second term on the right-hand-side is nothing other than the flow of volume associated with mass moving across a permeable boundary,
therefore the entire right-hand-side may be written as $\{\mathrm{d} V^\mathrm{II} / 
\mathrm{d} t\}_{T,\ p,\ \kappa}$.  Combining this with (\ref{eq:nu^2_k_dM^2_k_dt}) the
first term of (\ref{eq:sigma_tot_4}) becomes

\begin{equation}
	\frac{1}{T} \sum_{\alpha=\mathrm{II}}^{\mathrm{III}} \left( \Delta_{\mathrm{I} ,\ \alpha\ }p \right) \left\{ \frac{\mathrm{d} V^\alpha}{\mathrm{d} t}\right\}_{T,\ p,\ \kappa}.
	\label{eq:first_term}
\end{equation}

Turning attention to the second term of (\ref{eq:sigma_tot_4}), we notice that since interactions between lipids are neglected outside the membrane we are free to set
$\kappa = 0$ in all other regions, therefore only values of the summand for $\alpha = \mathrm{II}$ need be considered.  In a similar fashion to above, remembering
that $\Psi = \psi M = \sum_k \psi_k \rho_k$, we see that

\begin{equation}
	\begin{split}
		\sum_k \bar{\psi}_k^\mathrm{II} \frac{\mathrm{d} M_k^\mathrm{II}}{\mathrm{d} t} &= -\sum_k \int_\mathrm{II} \psi_k \rho_k \left( \boldsymbol{v}_k
		-\boldsymbol{v}^\mathrm{b} \right)\cdot\mathrm{d}\boldsymbol{A} \\
		&= \int_\mathrm{II} \Psi \boldsymbol{v}^\mathrm{b}\cdot\mathrm{d}\boldsymbol{A} - \sum_k \int_\mathrm{II} \psi_k\rho_k\boldsymbol{v}_k\cdot\mathrm{d}\boldsymbol{A}\\
		&= \frac{\mathrm{d} \Psi^\mathrm{II}}{\mathrm{d} t} - \int_\mathrm{II} \frac{\partial \Psi}{\partial t}\mathrm{d}V 
		- \sum_k\int_\mathrm{II} \psi_k\rho_k\boldsymbol{v}_k\cdot\mathrm{d}\boldsymbol{A}.
	\end{split}
	\label{eq:psi^2_k_dM^2_k_dt}
\end{equation}

\noindent Here, in analogy to above, we recognize the right-hand side as $\left\{ 
\mathrm{d} \Psi^\mathrm{II}/\mathrm{d} t \right\}_{T,\ p,\ \kappa}$, where the second term 
arises due to the fact that $\psi$ is not conserved: the curvature of the membrane can 
change spontaneously through the exchange of lipids between outer and inner monolayers.
Finally, assuming that the membrane thickness $l$ is small on the scale of the vesicle we 
may write

\begin{equation}
	\Psi^\mathrm{II} \equiv \int_\mathrm{II} \Psi \mathrm{d}V = l\int_\mathrm{m}\Psi\mathrm{d}A + \mathcal{O}(l^2),
	\label{eq:Psi^2}
\end{equation}

\noindent where, following from identifications (\ref{eq:mu_gamma_and_gamma}) the second 
term of (\ref{eq:sigma_tot_4}) may finally be written as

\begin{equation}
	-\frac{1}{T} \kappa \left\{ \frac{\mathrm{d} \Psi^\mathrm{II}}{\mathrm{d} t}\right\}_{T,\ p,\ \kappa} = -\frac{1}{T} \left\{ \frac{\mathrm{d} E_\mathrm{m}}{\mathrm{d} t}\right\}_{T,\ p,\ \kappa}.
	\label{eq:second_term}
\end{equation}

In analogy to the Gibbs-Duhem relation derived in Appendix II of \cite{NonEquilThermDeGrootMazur}, the third term of (\ref{eq:sigma_tot_4}) may be
simplified by writing

\begin{equation}
	\sum_k \bar{c}_k^\beta \{\Delta_{\alpha,\ \beta\ }\bar{\mu}_k\}_{T,\ p,\ \kappa \ } = 0,
	\label{eq:Gibbs_Duhem_like}
\end{equation} 

\noindent from which it can be seen that in the limit of dilute solutions, $c_l \ll c_w$

\begin{equation}
	\left\{\Delta_{\mathrm{II} ,\ \alpha\ }\bar{\mu}_w\right\}_{T,\ p,\ \kappa} = 0,\ \ \forall\ \alpha \in \{\mathrm{I},\ \mathrm{III}\}. \\
	\label{eq:Delta_mu_w=0}
\end{equation}

\noindent Furthermore assuming that the interior of the vesicle only contributes a 
negligible flow of lipids---that is, it is not
considered a reservoir---gives 

\begin{equation}
	\sum_{\alpha=\mathrm{II}}^{\mathrm{III}}\frac{\mathrm{d}M_l^\alpha}{\mathrm{d} t}\{\Delta_{\mathrm{I},\ \alpha\ }\bar{\mu}_l\}_{T,\ p,\ \kappa}
	= \frac{\mathrm{d}M_l^\mathrm{II}}{\mathrm{d} t}\{\Delta_{\mathrm{I} ,\ \mathrm{II}\ }\bar{\mu}_l\}_{T,\ p,\ \kappa}.
	\label{eq:dM_l^III_dt=0}
\end{equation}

\noindent Combining the results (\ref{eq:first_term}), (\ref{eq:second_term}) and 
(\ref{eq:dM_l^III_dt=0}) it is now possible to write (\ref{eq:sigma_tot_4}) as

\begin{equation}
\begin{split}
	T\sigma_{\mathrm{tot}} = & \sum_{\alpha=\mathrm{II}}^{\mathrm{III}}  
	\left( \Delta_{\mathrm{I},\ \alpha\ }p \right)\left\{\frac{\mathrm{d} V^\alpha}{\mathrm{d} t} \right\}_{T,\ p,\ \kappa} \\
	& \ \ -\left\{\frac{\mathrm{d} E_\mathrm{m}}{\mathrm{d} t} \right\}_{T,\ p,\ \kappa} + \frac{\mathrm{d}M_l^\mathrm{II}}{\mathrm{d} t} \{\Delta_{\mathrm{I},\ \mathrm{II}\ }\bar{\mu}_l\}_{T,\ p,\ \kappa}.
	\label{eq:sigma_tot_5}
\end{split}
\end{equation}

\noindent This expression can be further simplified by once again taking the membrane to be of constant thickness, $l$---very small on the scale of the vesicle---so that

\begin{equation}
	\left\{ \frac{\mathrm{d} V^{\mathrm{II}}}{\mathrm{d} t} \right\}_{T,\ p,\ \kappa} = l\left\{ \frac{\mathrm{d} A^\mathrm{m}}{\mathrm{d} t} \right\}_{T,\ p,\ \kappa} 
		+ \mathcal{O}(l^2),
	\label{eq:dV2_dt=ldA_dt}
\end{equation}

\noindent and

\begin{equation}
\begin{split}
	\left\{ \frac{\mathrm{d} V^{\mathrm{III}}}{\mathrm{d} t} \right\}_{T,\ p,\ \kappa} = & \left\{ \frac{\mathrm{d} V^{\mathrm{m}}}{\mathrm{d} t} \right\}_{T,\ p,\ \kappa} \\ 
	&- \frac{l}{2}\left\{ \frac{\mathrm{d} A^\mathrm{m}}{\mathrm{d} t} \right\}_{T,\ p,\ \kappa} + \mathcal{O}(l^2),
	\label{eq:dV3_dt=ldA_dt}
\end{split}
\end{equation}

\noindent where $A^\mathrm{m}$ and $V^\mathrm{m}$ are the area of, and volume enclosed by, 
the surface which bisects the membrane.  Finally, as $\kappa$ and $C_0$ are taken
to remain constant, it is assumed that, on average, every unit of mass (of lipids) 
accreted into the bilayer increases the area of the central bisecting surface by the same
factor, that is

\begin{equation}
	\frac{\mathrm{d} M^{\mathrm{II}}_l}{\mathrm{d} t} = a\left\{ \frac{\mathrm{d} A^{\mathrm{m}}}{\mathrm{d} t} \right\}_{T,\ p,\ \kappa},
	\label{eq:dM_dt=adA_dt}
\end{equation}

\noindent where $a$ is a constant.  This is in-line with the usual assumption that 
bilayers are essentially incompressible due to the separation of energy scales between 
stretching and bending energies~\cite{OFPP03}.  However, it is necessary to acknowledge 
that in certain circumstances (e.g.~highly compressed bilayers) thermal fluctuations are 
important~\cite{OFPP03}.  Applying the results, (\ref{eq:dV2_dt=ldA_dt}),  
(\ref{eq:dV3_dt=ldA_dt}) and (\ref{eq:dM_dt=adA_dt}) to (\ref{eq:sigma_tot_5}) gives

\begin{equation}
\begin{split}
	T\sigma_{\mathrm{tot}} =&  \left(\Delta_{\mathrm{I} ,\ \mathrm{III}\ }p\right)\left\{\frac{\mathrm{d} V^{\mathrm{m}}}{\mathrm{d} t}\right\}_{T,\ p,\ \kappa}  
	- \left\{\frac{\mathrm{d} E_m}{\mathrm{d} t} \right\}_{T,\ p,\ \kappa} \\
	& + \gamma\left\{\frac{\mathrm{d} A^{\mathrm{m}}}{\mathrm{d} t}\right\}_{T,\ p,\ \kappa},
	\label{eq:sigma_tot_7}
\end{split}
\end{equation}

\noindent where

\begin{equation}
	\gamma = \Big( \frac{l}{2}\left(p^\mathrm{III} + p^\mathrm{I}\right) - lp^\mathrm{II} + a\{\Delta_{\mathrm{I} ,\ \mathrm{II}\ }\bar{\mu}_l\}_{T,\ p,\ \kappa} \Big).
\end{equation}

\section{Perturbative expressions}\label{app:TenCalc}

In Section \ref{sec:pert} it is stated that geometrical quantities, $V$, $\xi_1$ and $\xi_2$ are required in terms of $R$, the radius of a sphere, and perturbation
$\epsilon (\theta)$.  This Appendix outlines how to arrive at the necessary results.   

Consider first $\xi_1 = \int H\mathrm{d}A = 2\pi\int_0^\pi H\sqrt{g}\mathrm{d}\theta$, where $H$ and $\sqrt{g}$ are given by (\ref{eq:H_axis}) and (\ref{eq:rootg_axis})
respectively.  Whilst it is possible to directly calculate a perturbative form for $\xi_1$--- by substituting (\ref{eq:rho_perturb}) and (\ref{eq:k_perturb}) into
(\ref{eq:H_axis}) and (\ref{eq:rootg_axis}) and integrating---here, a change of variable is introduced to simplify the manipulations slightly.  Motivated by the form of
(\ref{eq:rootg_axis}) introduce variables $\mathcal{R} (\theta)$ and $\Theta (\theta)$ such that

\begin{equation}
	\rho' = \mathcal{R}\cos{\Theta},\ \mathrm{and}\ \ k' = \mathcal{R}\sin{\Theta},
	\label{eq:RnewTheta}
\end{equation}

\noindent from which it follows that

\begin{equation}
	\mathcal{R} = \sqrt{ \left( \rho' \right)^2 + \left( k' \right)^2 },
	\label{eq:Rnew}
\end{equation}

\begin{equation}
	\Theta' = \frac{k''\rho' - \rho''k'}{\mathcal{R}^2},
	\label{eq:-Theta'}
\end{equation}

\noindent and

\begin{equation}
	\xi_1 = -\pi\int_0^\pi \mathrm{d}\theta \left( \rho\Theta' + k'\right)= \pi\int_0^\pi \mathrm{d}\theta \left( \rho'\Theta- k' \right),
	\label{eq:xi_1}
\end{equation}

\noindent where the second step of (\ref{eq:xi_1}) comes from integration by parts; noticing that $\rho(0) = \rho (\pi ) = 0$.  In order to find $\Theta$,
(\ref{eq:rho_perturb}) and (\ref{eq:k_perturb}) can be substituted into (\ref{eq:-Theta'}) giving

\begin{equation}
	\Theta' = -1 + \epsilon'' -\left(\epsilon\epsilon'\right)' + \frac{1}{3}\left(3\epsilon^2\epsilon' - \left(\epsilon'\right)^3\right)' + \mathcal{O} (\epsilon^4).
	\label{eq:-Theta'_pert}
\end{equation}

\noindent This expression can be easily integrated.  Applying the boundary conditions $\epsilon'\left(0\right)=\epsilon'\left(\pi\right)=0$ gives

\begin{equation}
	\Theta = -\theta + \epsilon' - \epsilon\epsilon' + \frac{1}{3}\left( 3\epsilon^2\epsilon' - \left(\epsilon'\right)^3\right) + \mathcal{O}\left(\epsilon^4\right).
	\label{eq-Theta_pert}
\end{equation}

\noindent Using (\ref{eq:rho_perturb}) and (\ref{eq:k_perturb}) to write down expressions for $\rho'$ and $k'$ and then substituting into (\ref{eq:xi_1}) along with the above gives

\begin{equation}
\begin{split}
	\xi_1 = \pi R\int_0^\pi\mathrm{d}\theta \sin{\theta} &\left[ 2 +2\epsilon +\left(\epsilon'\right)^2 + \left(\epsilon'\right)^2\left( \epsilon'' - \epsilon\right) \right.\\
	&\left. \ \ + \mathcal{O}\left(\epsilon^4\right)\right],
	\label{eq:xi_1_perturb}
\end{split}
\end{equation}

\noindent where the following result has been used

\begin{equation}
	\int_0^\pi \mathrm{d}\theta \cos{\theta} \left(\epsilon'\right)^3 = -3\int_0^\pi \mathrm{d}\theta \sin{\theta} \left(\epsilon'\right)^2\epsilon''.
	\label{eq:int_costheta_(e')^3}
\end{equation}

\noindent A similar procedure may now be applied to $\xi_2 = \int H^2\mathrm{d}A = 2\pi\int_0^\pi H^2\sqrt{g}\mathrm{d}\theta$. Using the same variable change as above

\begin{equation}
	\xi_2 = \frac{\pi}{2}\int_0^\pi \mathrm{d}\theta \left( \frac{\left(k'\right)^2}{\rho\mathcal{R}} + \frac{2 k'\Theta'}{\mathcal{R}} + \frac{\rho\left(\Theta'\right)^2}{\mathcal{R}}\right),
	\label{eq:xi_2}
\end{equation} 

\noindent it can be immediately seen from earlier definitions (\ref{eq:RnewTheta}) that the second term simplifies:

\begin{equation}
\begin{split}
	\int_0^\pi \mathrm{d}\theta \left(\frac{k'\Theta'}{\mathcal{R}}\right) &= \int_{\theta =0}^{\theta = \pi} \mathrm{d}\Theta \sin{\Theta} \\
	&= -\left[\cos{\Theta}\right]_{\theta=0}^{\theta=\pi} = -\left[\frac{\rho'}{\mathcal{R}}\right]_{\theta=0}^{\theta=\pi} = 2,
\label{eq:xi_2_simplification}
\end{split}
\end{equation}

\noindent where the last step follows from the definitions of $\rho'$, $\mathcal{R}$ and boundary conditions $\epsilon'\left(0\right)=\epsilon'\left(\pi\right)=0$.  The
remaining terms of (\ref{eq:xi_2}) can be calculated in a straightforward way though the lengthy intermediate steps have been omitted here.  The result is that

\begin{equation}
\begin{split}
	\xi_2 = \pi\int_0^\pi \mathrm{d}\theta\sin{\theta}&\left[2 - \left(\epsilon'\cot{\theta} +\epsilon''\right) -\left(\epsilon'\right)^2 \right.\\
	&\ \  + \frac{1}{2}\left(\epsilon'\cot{\theta} +\epsilon''\right)^2 + 2\left(\epsilon'\right)^2\left(\epsilon + \epsilon''\right) \\
	&\ \ \left. -  \epsilon\left(\epsilon'\cot{\theta} +\epsilon''\right)^2 + \mathcal{O}\left(\epsilon^4\right) \right].
	\label{eq:xi_2_perturb}
\end{split}
\end{equation}

\noindent In the same fashion as shown in Section \ref{sec:pert} for the surface area, it is possible to re-write (\ref{eq:xi_1_perturb}) and (\ref{eq:xi_2_perturb}) in
terms of the operator $\hat{L}^2$---defined in (\ref{eq:L^2_theta})---using integration by parts.  

\begin{equation}
\begin{split}
	\xi_1 = \pi R\int_0^{\pi}\mathrm{d}\theta \sin\theta&\left[ 2 + 2\epsilon - \epsilon\hat{L}^2\epsilon + \frac{1}{2}\epsilon^2\hat{L}^2\epsilon\right. \\
	& \ \ \left. - \frac{1}{2} (\epsilon ')^2\hat{L}^2\epsilon + \mathcal{O} (\epsilon^4)\right],
	\label{eq:xi_1_L^2}
\end{split}
\end{equation}

\noindent and

\begin{equation}
\begin{split}
	\xi_2 = \pi\int_0^{\pi}\mathrm{d}\theta\sin\theta&\left[2 - \hat{L}^2\epsilon + \epsilon\hat{L}^2\epsilon + \frac{1}{2}\left(\hat{L}^2\epsilon\right)^2 - \epsilon^2\hat{L}^2\epsilon \right. \\
	 & \ \ \left. - \epsilon\left(\hat{L}^2\epsilon\right)^2 - (\epsilon ')^2\hat{L}^2\epsilon + \mathcal{O} (\epsilon^4) \right].
	\label{eq:xi_2_L^2}
\end{split}
\end{equation}

\noindent These expressions can then be integrated using (\ref{eq:sum_a_l_Y_l}), (\ref{eq:zonal_orthogonal}) and (\ref{eq:L^2_Y_l}) to give the following results:

\begin{equation}
\begin{split}
	\xi_1 =& 4\pi R + \varepsilon^2\frac{R}{2} \sum_{l=2}^\infty a_l^2 l(l+1) \\ 
	& + \varepsilon^3\frac{R}{8} \sum_{l_1, l_2, l_3}^\infty a_{l_1} a_{l_2} a_{l_3} \bigg\{l_3 (l_3 + 1)\Big[2l_2 (l_2 +1) \\
	&  -l_3 (l_3+1) - 2\Big]\bigg\} f(l_1,\ l_2,\ l_3) + \mathcal{O}(\varepsilon^4),
	\label{eq:xi_1_int}
\end{split}
\end{equation}

\noindent and

\begin{equation}
\begin{split}
	\xi_2 =& 4\pi + \varepsilon^2\frac{1}{4} \sum_{l=2}^\infty a_l^2 \bigg\{l(l+1)\Big[l(l+1) - 2\Big]\bigg\} \\ 
	& - \varepsilon^3\frac{1}{4} \sum_{l_1, l_2, l_3}^\infty  
	a_{l_1} a_{l_2} a_{l_3} \bigg\{l_3 (l_3 + 1)\Big[l_3 (l_3 + 1) - 2\Big]\bigg\} \\
	&\times f(l_1,\ l_2,\ l_3) + \mathcal{O} (\varepsilon^4),
	\label{eq:xi_2_int}
\end{split}
\end{equation}

\noindent where, the function $f(l_1,\ l_2,\ l_3)$ is given by 

\begin{equation}
\begin{split}
f(l_1,\ l_2,\ l_3) &= 2\pi\int_0^{\pi} \mathrm{d}\theta\sin\theta\ Y_{l_1} (\theta) Y_{l_2} (\theta) Y_{l_3} (\theta) \\
&= \sqrt{\frac{(2l_1 + 1)(2l_2 + 1)(2l_3 + 1)}{4\pi}} \\
&\quad\times\left(\begin{array}{ccc}
l_1 & l_2 & l_3 \\
0 & 0 & 0
\end{array}\right)^2,
\end{split}
\end{equation}

\noindent and

\begin{equation}
\left(\begin{array}{ccc}
l_1 & l_2 & l_3 \\
0 & 0 & 0
\end{array}\right)
\end{equation}

\noindent is a Wigner 3-j symbol (see, for example, Appendix C.I of~\cite{Messiah}) with $m$-values set to zero.  The symbol is zero unless the triangle condition, $\vert
l_1 - l_2 \vert \leq l_3 \leq l_1 + l_2$, holds.  Finally, the volume contained by an axisymmetric surface is given by

\begin{eqnarray}
	V &=& 2\pi \int_0^R \mathrm{d} R \int_0^\pi \mathrm{d}\theta \left(\frac{\partial \boldsymbol{r}}{\partial R}\right)\cdot\hat{\boldsymbol{n}}\sqrt{g} \nonumber\\
	&=&\frac{2}{3}\pi R^3 \int_0^\pi\sin{\theta}\mathrm{d}\theta\left[1+3\epsilon + 3\epsilon^2 + \epsilon^3\right],
	\label{eq:V}
\end{eqnarray}

\noindent which can also be integrated using the properties of the zonal harmonics to give Eq.~(\ref{eq:V_int_duplicate}) in the main text. 

\section{Partial derivatives}\label{app:stability}

In order to write down the right-hand side of (\ref{eq:constitutive}), partial derivatives (\ref{eq:dEm_dV_A}) and (\ref{eq:dEm_dA_V}) are needed in terms of $r$, the radius
of a sphere with equivalent surface area, and $\varepsilon$.  This Appendix provides the details of the calculation.  

First, after invoking the growth law (\ref{eq:growth}), the undeformed radius $R$ appearing in the expressions for $\xi_1$ and $V$---(\ref{eq:xi_1_int}) and
(\ref{eq:V_int_duplicate}) respectively---must be eliminated in favor of $r$.  Using the relation (\ref{eq:R(r)}) gives   

\begin{equation}
\begin{split}
	\xi_1 =& 4\pi r - \varepsilon^2\frac{r}{4} \sum_{l=2}^\infty a_l^2 \left[2 - l(l+1)\right] \\ 
	& + \varepsilon^3\frac{r}{8} \sum_{l_1, l_2, l_3}^\infty a_{l_1} a_{l_2} a_{l_3} \bigg\{l_3 (l_3 + 1)\Big[2l_2 (l_2 +1) \\
	&  -l_3 (l_3+1) - 2\Big]\bigg\} f(l_1,\ l_2,\ l_3) + \mathcal{O}(\varepsilon^4),
	\label{eq:xi_1_int_r}
\end{split}
\end{equation}

\noindent and

\begin{equation}
\begin{split}
	V =& \frac{4}{3}\pi r^3 + \varepsilon^2 \frac{r^3}{4} \sum_{l=2}^\infty a_l^2 \left[2 - l(l+1)\right] \\
	&+ \varepsilon^3\frac{r^3}{3}\sum_{l_1, l_2, l_3}^\infty a_{l_1} a_{l_2} a_{l_3} f(l_1,\ l_2,\ l_3) + \mathcal{O}(\varepsilon^4),
	\label{eq:V_int_r}
\end{split}
\end{equation}

\noindent where $\xi_2$ remains unchanged.  We may now follow~\cite{VesRGMDFAJM} and introduce the reduced volume $v$, such that  

\begin{equation}
	V = \frac{4\pi r^3}{3} v.
\end{equation}

\noindent Noticing that $v$ is a function of $\varepsilon$ only, implies that

\begin{equation}
	\left( \frac{\partial \varepsilon}{\partial V} \right)_{r} = \frac{1}{4\pi r^3/3} [v'(\varepsilon)]^{-1},
	\label{eq:d_epsilon_dV_r}
\end{equation}

\noindent and

\begin{equation}
	\left( \frac{\partial \varepsilon}{\partial r} \right)_{V} = -\frac{3}{r}\,\frac{V}{4\pi r^3/3} [v'(\varepsilon)]^{-1},
	\label{eq:d_epsilon_dr_V}
\end{equation}

\noindent where using the notation of (\ref{eq:xi_1_simple})-(\ref{eq:V_simple})  

\begin{equation}
	v'(\varepsilon) = \frac{\mathrm{d} v}{\mathrm{d} \varepsilon} = 2\varepsilon V^{(2)} + 3\varepsilon^2 V^{(3)} +\mathcal{O}(\varepsilon^3).
\end{equation}

\noindent Substituting (\ref{eq:d_epsilon_dV_r}) into (\ref{eq:dEm_dV_A}) and (\ref{eq:d_epsilon_dr_V}) into (\ref{eq:dEm_dA_V}) gives

\begin{equation}
	\left( \frac{\partial E_{\mathrm{m}}}{\partial V} \right)_{A} = \frac{\left[ v'(\varepsilon) \right]^{-1}}{4\pi r^{3}/3}\,\left( \frac{\partial E_{m}}{\partial \varepsilon} \right)_{r},
	\label{eq:dEm_dV_A_2}
\end{equation}

\noindent and

\begin{equation}
	\left( \frac{\partial E_{\mathrm{m}}}{\partial A} \right)_{V} = \frac{1}{8\pi r}\,\left( \frac{\partial E_{\mathrm{m}}}{\partial r} \right)_{\varepsilon}
		- \frac{3V}{2A}\,\left( \frac{\partial E_{\mathrm{m}}}{\partial V} \right)_{A}.
	\label{eq:dEm_dA_V_2}
\end{equation}

\noindent Writing $E_\mathrm{m} = 2\kappa\xi_2 - 2\kappa C_0 \xi_1 + \kappa C_0^2 A / 2$, the partial derivative $\left(\partial E_\mathrm{m} / \partial \varepsilon\right)_r$
can be found directly using (\ref{eq:xi_1_simple}), (\ref{eq:xi_2_simple}) and $A=4\pi r^2$.  Substituting into (\ref{eq:dEm_dV_A_2}) gives

\begin{equation}
\begin{split}
	\left( \frac{\partial E_\mathrm{m}}{\partial V} \right)_A =& \frac{6\kappa\xi_2^{(2)}}{r^3 V^{(2)}} - \frac{6\kappa C_0\xi_1^{(2)}}{r^2 V^{(2)}}  + \varepsilon \frac{9\kappa}{r^3 V^{(2)}}\\
	&\times\left[ \left( \xi_2^{(3)} - \frac{V^{(3)}\xi_2^{(2)}}{V^{(2)}} \right) - C_0 r\left( \xi_1^{(3)} + \frac{V^{(3)}}{3} \right)\right],
	\label{eq:dEm_dV_A_pert}
\end{split}
\end{equation}

\noindent where the fact that $\xi_1^{(2)}/V^{(2)} = -1/3$ has been used.  Similarly, $\left(\partial E_\mathrm{m} / \partial r \right)_\varepsilon$ can also be found from
(\ref{eq:xi_1_simple}), (\ref{eq:xi_2_simple}) and $A=4\pi r^2$.  Substituting this result and (\ref{eq:dEm_dV_A_pert}) into (\ref{eq:dEm_dA_V_2}) gives 

\begin{equation}
\begin{split}
	\left( \frac{\partial E_\mathrm{m}}{\partial A} \right)_V =& -2\frac{\kappa C_0}{r} + \frac{\kappa C_0^2}{2} - \frac{3\kappa\xi_2^{(2)}}{r^2 V^{(2)}} \\
	& - \varepsilon\frac{9\kappa}{2 r^2 V^{(2)}} \left[ \left( \xi_2^{(3)} - \frac{V^{(3)}\xi_2^{(2)}}{V^{(2)}} \right)\right. \\
	&\quad\quad\quad\quad\quad\quad \left. - C_0 r\left( \xi_1^{(3)} + \frac{V^{(3)}}{3} \right)\right],
\end{split}
	\label{eq:dEm_dA_V_pert}
\end{equation}

\noindent where once again $\xi_1^{(2)}/V^{(2)} = -1/3$ has been used.


\begin{thebibliography}{44}
\expandafter\ifx\csname natexlab\endcsname\relax\def\natexlab#1{#1}\fi
\expandafter\ifx\csname bibnamefont\endcsname\relax
  \def\bibnamefont#1{#1}\fi
\expandafter\ifx\csname bibfnamefont\endcsname\relax
  \def\bibfnamefont#1{#1}\fi
\expandafter\ifx\csname citenamefont\endcsname\relax
  \def\citenamefont#1{#1}\fi
\expandafter\ifx\csname url\endcsname\relax
  \def\url#1{\texttt{#1}}\fi
\expandafter\ifx\csname urlprefix\endcsname\relax\def\urlprefix{URL }\fi
\providecommand{\bibinfo}[2]{#2}
\providecommand{\eprint}[2][]{\url{#2}}

\bibitem[{\citenamefont{{U. Seifert}}(1997)}]{VesReviewSeifert}
\bibinfo{author}{\bibnamefont{{U. Seifert}}}, \bibinfo{journal}{Adv. Phys.}
  \textbf{\bibinfo{volume}{46}}, \bibinfo{pages}{13} (\bibinfo{year}{1997}).

\bibitem[{\citenamefont{{Z.-C. Ou-Yang} et~al.}(1999)\citenamefont{{Z.-C.
  Ou-Yang}, {J.-X. Liu}, and {Y.-Z. Xie}}}]{VesYangLiuXie}
\bibinfo{author}{\bibnamefont{{Z.-C. Ou-Yang}}},
  \bibinfo{author}{\bibnamefont{{J.-X. Liu}}}, \bibnamefont{and}
  \bibinfo{author}{\bibnamefont{{Y.-Z. Xie}}}, \emph{\bibinfo{title}{Geometric
  {M}ethods in the {E}lastic {T}heory of {M}embranes in {L}iquid {C}rystal
  {P}hases}} (\bibinfo{publisher}{World Scientific Publishing},
  \bibinfo{address}{Singapore}, \bibinfo{year}{1999}).

\bibitem[{\citenamefont{{R. Lipowsky}}(1991)}]{LipowskyNature}
\bibinfo{author}{\bibnamefont{{R. Lipowsky}}}, \bibinfo{journal}{Nature}
  \textbf{\bibinfo{volume}{349}}, \bibinfo{pages}{475} (\bibinfo{year}{1991}).

\bibitem[{\citenamefont{{P. L. Luisi}}(2006)}]{LifeLuisi}
\bibinfo{author}{\bibnamefont{{P. L. Luisi}}}, \emph{\bibinfo{title}{The
  {E}mergence of {L}ife}} (\bibinfo{publisher}{Chapter 10. Cambridge University
  Press}, \bibinfo{address}{Cambridge}, \bibinfo{year}{2006}).

\bibitem[{\citenamefont{{P. B. Canham}}(1970)}]{CanhamJTB}
\bibinfo{author}{\bibnamefont{{P. B. Canham}}}, \bibinfo{journal}{J. Theor.
  Biol.} \textbf{\bibinfo{volume}{26}}, \bibinfo{pages}{61}
  (\bibinfo{year}{1970}).

\bibitem[{\citenamefont{{W. Helfrich}}(1973)}]{VesHelfrich}
\bibinfo{author}{\bibnamefont{{W. Helfrich}}}, \bibinfo{journal}{Z.
  Naturforsch.} \textbf{\bibinfo{volume}{28}}, \bibinfo{pages}{693}
  (\bibinfo{year}{1973}).

\bibitem[{\citenamefont{{E. A. Evans}}(1974)}]{EvansJBiophys}
\bibinfo{author}{\bibnamefont{{E. A. Evans}}}, \bibinfo{journal}{J. Biophys}
  \textbf{\bibinfo{volume}{14}}, \bibinfo{pages}{923} (\bibinfo{year}{1974}).

\bibitem[{\citenamefont{{B. Bo\u{z}i\u{c}} and {S.
  Svetina}}(2004)}]{SvetinaBozicBiophys}
\bibinfo{author}{\bibnamefont{{B. Bo\u{z}i\u{c}}}} \bibnamefont{and}
  \bibinfo{author}{\bibnamefont{{S. Svetina}}}, \bibinfo{journal}{Eur. Biophys.
  J.} \textbf{\bibinfo{volume}{33}}, \bibinfo{pages}{565}
  (\bibinfo{year}{2004}).

\bibitem[{\citenamefont{{B. Bo\u{z}i\u{c}} and {S.
  Svetina}}(2007)}]{SvetinaBozicEurPhysJE}
\bibinfo{author}{\bibnamefont{{B. Bo\u{z}i\u{c}}}} \bibnamefont{and}
  \bibinfo{author}{\bibnamefont{{S. Svetina}}}, \bibinfo{journal}{Eur. Phys. J.
  E} \textbf{\bibinfo{volume}{24}}, \bibinfo{pages}{79} (\bibinfo{year}{2007}).

\bibitem[{\citenamefont{{R. V. Sol\'e} and {J.
  Mac\'{\i}a}}(2007)}]{SoleMaciaJTheorBiol}
\bibinfo{author}{\bibnamefont{{R. V. Sol\'e}}} \bibnamefont{and}
  \bibinfo{author}{\bibnamefont{{J. Mac\'{\i}a}}}, \bibinfo{journal}{J. Theor.
  Biol.} \textbf{\bibinfo{volume}{245}}, \bibinfo{pages}{400}
  (\bibinfo{year}{2007}).

\bibitem[{\citenamefont{{R. V. Sol\'e} et~al.}(2007)\citenamefont{{R. V.
  Sol\'e}, {A. Munteanu}, {C. Rodriguez-Caso}, and {J.
  Mac\'{\i}a}}}]{SoleMaciaPhilTransRSB}
\bibinfo{author}{\bibnamefont{{R. V. Sol\'e}}},
  \bibinfo{author}{\bibnamefont{{A. Munteanu}}},
  \bibinfo{author}{\bibnamefont{{C. Rodriguez-Caso}}}, \bibnamefont{and}
  \bibinfo{author}{\bibnamefont{{J. Mac\'{\i}a}}}, \bibinfo{journal}{Phil.
  Trans. R. Soc B} \textbf{\bibinfo{volume}{362}}, \bibinfo{pages}{1821}
  (\bibinfo{year}{2007}).

\bibitem[{\citenamefont{{D. Fanelli} and {A. J. McKane}}(2008)}]{VesAJM}
\bibinfo{author}{\bibnamefont{{D. Fanelli}}} \bibnamefont{and}
  \bibinfo{author}{\bibnamefont{{A. J. McKane}}}, \bibinfo{journal}{Phys. Rev.
  E} \textbf{\bibinfo{volume}{78}}, \bibinfo{pages}{051406}
  (\bibinfo{year}{2008}).

\bibitem[{\citenamefont{{B. Bo\u{z}i\u{c}} and {S.
  Svetina}}(2009)}]{BozicSvetinaComment}
\bibinfo{author}{\bibnamefont{{B. Bo\u{z}i\u{c}}}} \bibnamefont{and}
  \bibinfo{author}{\bibnamefont{{S. Svetina}}}, \bibinfo{journal}{Phys. Rev. E}
  \textbf{\bibinfo{volume}{80}}, \bibinfo{pages}{013401}
  (\bibinfo{year}{2009}).

\bibitem[{\citenamefont{{D. Fanelli} and {A. J.
  McKane}}(2009)}]{FanelliMckaneReplyComment}
\bibinfo{author}{\bibnamefont{{D. Fanelli}}} \bibnamefont{and}
  \bibinfo{author}{\bibnamefont{{A. J. McKane}}}, \bibinfo{journal}{Phys. Rev.
  E} \textbf{\bibinfo{volume}{80}}, \bibinfo{pages}{013402}
  (\bibinfo{year}{2009}).

\bibitem[{\citenamefont{{R. G. Morris} et~al.}(2010)\citenamefont{{R. G.
  Morris}, {D. Fanelli}, and {A. J. McKane}}}]{VesRGMDFAJM}
\bibinfo{author}{\bibnamefont{{R. G. Morris}}},
  \bibinfo{author}{\bibnamefont{{D. Fanelli}}}, \bibnamefont{and}
  \bibinfo{author}{\bibnamefont{{A. J. McKane}}}, \bibinfo{journal}{Phys. Rev.
  E} \textbf{\bibinfo{volume}{82}}, \bibinfo{pages}{031125}
  (\bibinfo{year}{2010}).

\bibitem[{\citenamefont{{S. R. de Groot} and {P.
  Mazur}}(1984)}]{NonEquilThermDeGrootMazur}
\bibinfo{author}{\bibnamefont{{S. R. de Groot}}} \bibnamefont{and}
  \bibinfo{author}{\bibnamefont{{P. Mazur}}},
  \emph{\bibinfo{title}{Non-equilibrium {T}hermodynamics}}
  (\bibinfo{publisher}{Dover Publications}, \bibinfo{address}{New York},
  \bibinfo{year}{1984}).

\bibitem[{\citenamefont{{M. Wortis} et~al.}(1997)\citenamefont{{M. Wortis}, {M.
  Jari\'{c}}, and {U. Seifert}}}]{MWMJ97}
\bibinfo{author}{\bibnamefont{{M. Wortis}}}, \bibinfo{author}{\bibnamefont{{M.
  Jari\'{c}}}}, \bibnamefont{and} \bibinfo{author}{\bibnamefont{{U. Seifert}}},
  \bibinfo{journal}{J. Mol. Liq.} \textbf{\bibinfo{volume}{71}},
  \bibinfo{pages}{195} (\bibinfo{year}{1997}).

\bibitem[{\citenamefont{{W. Helfrich} and {R.-M. Servuss}}(1984)}]{WHRS84}
\bibinfo{author}{\bibnamefont{{W. Helfrich}}} \bibnamefont{and}
  \bibinfo{author}{\bibnamefont{{R.-M. Servuss}}}, \bibinfo{journal}{Nuovo
  Cimento D} \textbf{\bibinfo{volume}{3}}, \bibinfo{pages}{137}
  (\bibinfo{year}{1984}).

\bibitem[{\citenamefont{{U. Seifert}}(1995)}]{US95}
\bibinfo{author}{\bibnamefont{{U. Seifert}}}, \bibinfo{journal}{Z. Phys. B:
  Condens. Matter} \textbf{\bibinfo{volume}{97}}, \bibinfo{pages}{299}
  (\bibinfo{year}{1995}).

\bibitem[{\citenamefont{{V. Heinrich} et~al.}(1997)\citenamefont{{V. Heinrich},
  {F. Sev\u{s}ek}, {S. Svetina}, and {B. \u{Z}ek\u{s}}}}]{VHFS+97}
\bibinfo{author}{\bibnamefont{{V. Heinrich}}},
  \bibinfo{author}{\bibnamefont{{F. Sev\u{s}ek}}},
  \bibinfo{author}{\bibnamefont{{S. Svetina}}}, \bibnamefont{and}
  \bibinfo{author}{\bibnamefont{{B. \u{Z}ek\u{s}}}}, \bibinfo{journal}{Phys.
  Rev. E} \textbf{\bibinfo{volume}{55}}, \bibinfo{pages}{1809}
  (\bibinfo{year}{1997}).

\bibitem[{\citenamefont{{O. Farago} and {P. Pincus}}(2003)}]{OFPP03}
\bibinfo{author}{\bibnamefont{{O. Farago}}} \bibnamefont{and}
  \bibinfo{author}{\bibnamefont{{P. Pincus}}}, \bibinfo{journal}{Eur. Phys. J.
  E} \textbf{\bibinfo{volume}{11}}, \bibinfo{pages}{399}
  (\bibinfo{year}{2003}).

\bibitem[{\citenamefont{{F. C. Frank}}(1958)}]{LiquidCrystalEnergyDensityFrank}
\bibinfo{author}{\bibnamefont{{F. C. Frank}}}, \bibinfo{journal}{Discuss.
  Faraday Soc.} \textbf{\bibinfo{volume}{25}}, \bibinfo{pages}{19}
  (\bibinfo{year}{1958}).

\bibitem[{\citenamefont{{J. L.
  Ericksen}}(1961)}]{ConservationLiquidCrystalsEricksen}
\bibinfo{author}{\bibnamefont{{J. L. Ericksen}}}, \bibinfo{journal}{J. Rheol.}
  \textbf{\bibinfo{volume}{5}}, \bibinfo{pages}{23} (\bibinfo{year}{1961}).

\bibitem[{\citenamefont{{F. M.
  Leslie}}(1968)}]{ConstitutiveEqnsLiquidCrystalsLeslie}
\bibinfo{author}{\bibnamefont{{F. M. Leslie}}}, \bibinfo{journal}{Arch. Ration.
  Mech. Anal} \textbf{\bibinfo{volume}{28}}, \bibinfo{pages}{265}
  (\bibinfo{year}{1968}).

\bibitem[{\citenamefont{{H.-W. Huang}}(1971)}]{NematohydrodynamicsHuang}
\bibinfo{author}{\bibnamefont{{H.-W. Huang}}}, \bibinfo{journal}{Phys. Rev.
  Lett.} \textbf{\bibinfo{volume}{26}}, \bibinfo{pages}{1525}
  (\bibinfo{year}{1971}).

\bibitem[{\citenamefont{{A. J. Staverman}}(1952)}]{Staverman}
\bibinfo{author}{\bibnamefont{{A. J. Staverman}}}, \bibinfo{journal}{Trans.
  Faraday Soc.} \textbf{\bibinfo{volume}{48}}, \bibinfo{pages}{176}
  (\bibinfo{year}{1952}).

\bibitem[{\citenamefont{{S. R. De Groot} et~al.}(1966)\citenamefont{{S. R. De
  Groot}, {P. Mazur}, and {A. Michels}}}]{DeGrootMazurMichels}
\bibinfo{author}{\bibnamefont{{S. R. De Groot}}},
  \bibinfo{author}{\bibnamefont{{P. Mazur}}}, \bibnamefont{and}
  \bibinfo{author}{\bibnamefont{{A. Michels}}}, \bibinfo{journal}{Appl. sci.
  Res.} \textbf{\bibinfo{volume}{15}}, \bibinfo{pages}{261}
  (\bibinfo{year}{1966}).

\bibitem[{\citenamefont{{J. W. Lorimer}}(1985)}]{LorimerMembrane}
\bibinfo{author}{\bibnamefont{{J. W. Lorimer}}}, \bibinfo{journal}{J. Membrane
  Sci.} \textbf{\bibinfo{volume}{25}}, \bibinfo{pages}{181}
  (\bibinfo{year}{1985}).

\bibitem[{\citenamefont{{B. Baranowski}}(1990)}]{BaranowskiMembrane}
\bibinfo{author}{\bibnamefont{{B. Baranowski}}}, \bibinfo{journal}{J. Membrane
  Sci.} \textbf{\bibinfo{volume}{57}}, \bibinfo{pages}{119}
  (\bibinfo{year}{1990}).

\bibitem[{\citenamefont{{C. Tanford}}(1973)}]{HydrophobicEffect}
\bibinfo{author}{\bibnamefont{{C. Tanford}}}, \emph{\bibinfo{title}{The
  {H}ydrophobic {E}ffect}} (\bibinfo{publisher}{John Wiley \& Sons},
  \bibinfo{address}{New York}, \bibinfo{year}{1973}).

\bibitem[{\citenamefont{{J. N. Israelachvili} et~al.}(1976)\citenamefont{{J. N.
  Israelachvili}, {D. J. Mitchell}, and {B. W. Ninham}}}]{SelfAssemblyIsrael}
\bibinfo{author}{\bibnamefont{{J. N. Israelachvili}}},
  \bibinfo{author}{\bibnamefont{{D. J. Mitchell}}}, \bibnamefont{and}
  \bibinfo{author}{\bibnamefont{{B. W. Ninham}}}, \bibinfo{journal}{J. Chem.
  Soc., Faraday Trans. 2} \textbf{\bibinfo{volume}{72}}, \bibinfo{pages}{1525}
  (\bibinfo{year}{1976}).

\bibitem[{\citenamefont{{O. H. Samuli Ollila} et~al.}(2009)}]{OHSO+09}
\bibinfo{author}{\bibnamefont{{O. H. Samuli Ollila}}},
	\bibinfo{author}{\bibnamefont{{H. J. Risselada}}},
	\bibinfo{author}{\bibnamefont{{M. Louhivuori}}},
	\bibinfo{author}{\bibnamefont{{E. Lindahl}}},
\bibnamefont{and}
  \bibinfo{author}{\bibnamefont{{S. J. Marrink}}}, \bibinfo{journal}{Phys. Rev. Lett.} 
  \textbf{\bibinfo{volume}{102}}, \bibinfo{pages}{078101}
  (\bibinfo{year}{2009}).

\bibitem[{\citenamefont{{A. G. Petrov} and {I. Bivas}}(1984)}]{PetrovBivas}
\bibinfo{author}{\bibnamefont{{A. G. Petrov}}} \bibnamefont{and}
  \bibinfo{author}{\bibnamefont{{I. Bivas}}}, \bibinfo{journal}{Progr. Surf.
  Sci.} \textbf{\bibinfo{volume}{16}}, \bibinfo{pages}{398}
  (\bibinfo{year}{1984}).

\bibitem[{\citenamefont{{O. Kedem} and {A.
  Katchalsky}}(1958)}]{KedemKatchalBiophys}
\bibinfo{author}{\bibnamefont{{O. Kedem}}} \bibnamefont{and}
  \bibinfo{author}{\bibnamefont{{A. Katchalsky}}}, \bibinfo{journal}{Biochim.
  Biophys. Acta} \textbf{\bibinfo{volume}{27}}, \bibinfo{pages}{229}
  (\bibinfo{year}{1958}).

\bibitem[{\citenamefont{{O. Kedem} and {A.
  Katchalsky}}(1963)}]{KedemKatchalFaraday}
\bibinfo{author}{\bibnamefont{{O. Kedem}}} \bibnamefont{and}
  \bibinfo{author}{\bibnamefont{{A. Katchalsky}}}, \bibinfo{journal}{Trans.
  Faraday Soc.} \textbf{\bibinfo{volume}{59}}, \bibinfo{pages}{1918}
  (\bibinfo{year}{1963}).

\bibitem[{\citenamefont{{D. C. Kay}}(1998)}]{SchaumTenCalc}
\bibinfo{author}{\bibnamefont{{D. C. Kay}}}, \emph{\bibinfo{title}{Schaum's
  {O}utline of {T}heory and {P}roblems of {T}ensor {C}alculus}}
  (\bibinfo{publisher}{McGraw-Hill}, \bibinfo{address}{New York},
  \bibinfo{year}{1998}).

\bibitem[{\citenamefont{{S. A. Safran}}(1983)}]{VesSaf}
\bibinfo{author}{\bibnamefont{{S. A. Safran}}}, \bibinfo{journal}{J. Chem.
  Phys.} \textbf{\bibinfo{volume}{78}}, \bibinfo{pages}{2073}
  (\bibinfo{year}{1983}).

\bibitem[{\citenamefont{{G. B. Arfken}}(1985)}]{Arfken}
\bibinfo{author}{\bibnamefont{{G. B. Arfken}}},
  \emph{\bibinfo{title}{Mathematical {M}ethods for {P}hysicists}}
  (\bibinfo{publisher}{Elsevier}, \bibinfo{address}{Amsterdam},
  \bibinfo{year}{1985}).

\bibitem[{\citenamefont{{D. Marsh}}(2006)}]{Marsh2006}
\bibinfo{author}{\bibnamefont{{D. Marsh}}}, \bibinfo{journal}{Chem. Phys.
  Lipids} \textbf{\bibinfo{volume}{144}}, \bibinfo{pages}{146}
  (\bibinfo{year}{2006}).

\bibitem[{\citenamefont{{D. W. R. Gruen} and {J.
  Wolfe}}(1982)}]{LateralTensionGruen}
\bibinfo{author}{\bibnamefont{{D. W. R. Gruen}}} \bibnamefont{and}
  \bibinfo{author}{\bibnamefont{{J. Wolfe}}}, \bibinfo{journal}{Biochim.
  Biophys. Acta} \textbf{\bibinfo{volume}{2}}, \bibinfo{pages}{572}
  (\bibinfo{year}{1982}).

\bibitem[{\citenamefont{{J. K\"{a}s} and {E. Sackmann}}(1991)}]{JKES91}
\bibinfo{author}{\bibnamefont{{J. K\"{a}s}}} \bibnamefont{and}
  \bibinfo{author}{\bibnamefont{{E. Sackmann}}}, \bibinfo{journal}{Biophys. J.}
  \textbf{\bibinfo{volume}{60}}, \bibinfo{pages}{825} (\bibinfo{year}{1991}).

\bibitem[{\citenamefont{{J. Pencer} et~al.}(2001)\citenamefont{{J. Pencer}, {G.
  F.~White}, and {F. R.~Hallett}}}]{Pen01}
\bibinfo{author}{\bibnamefont{{J. Pencer}}}, \bibinfo{author}{\bibnamefont{{G.
  F.~White}}}, \bibnamefont{and} \bibinfo{author}{\bibnamefont{{F.
  R.~Hallett}}}, \bibinfo{journal}{Biophys. J.} \textbf{\bibinfo{volume}{81}},
  \bibinfo{pages}{2716} (\bibinfo{year}{2001}).

\bibitem[{\citenamefont{{K. Berndl} et~al.}(1990)\citenamefont{{K. Berndl}, {J.
  K\"{a}s}, {R. Lipowsky}, and {E. Sackmann}}}]{KBJK+90}
\bibinfo{author}{\bibnamefont{{K. Berndl}}}, \bibinfo{author}{\bibnamefont{{J.
  K\"{a}s}}}, \bibinfo{author}{\bibnamefont{{R. Lipowsky}}}, \bibnamefont{and}
  \bibinfo{author}{\bibnamefont{{E. Sackmann}}}, \bibinfo{journal}{Europhys.
  Lett.} \textbf{\bibinfo{volume}{13}}, \bibinfo{pages}{659}
  (\bibinfo{year}{1990}).

\bibitem[{\citenamefont{{L. Miao} et~al.}(1994)\citenamefont{{L. Miao}, {U.
  Seifert}, {M. Wortis}, and {H-G. D\"{o}bereiner}}}]{miao94}
\bibinfo{author}{\bibnamefont{{L. Miao}}}, \bibinfo{author}{\bibnamefont{{U.
  Seifert}}}, \bibinfo{author}{\bibnamefont{{M. Wortis}}}, \bibnamefont{and}
  \bibinfo{author}{\bibnamefont{{H-G. D\"{o}bereiner}}},
  \bibinfo{journal}{Phys. Rev. E} \textbf{\bibinfo{volume}{49}},
  \bibinfo{pages}{5389} (\bibinfo{year}{1994}).

\bibitem[{\citenamefont{{A. Messiah}}(1962)}]{Messiah}
\bibinfo{author}{\bibnamefont{{A. Messiah}}}, \emph{\bibinfo{title}{Quantum
  {M}echanics {V}ol. 2}} (\bibinfo{publisher}{North-Holland},
  \bibinfo{address}{Amsterdam, Netherlands}, \bibinfo{year}{1962}).

\end{thebibliography}
\end{document}